\documentclass[preprint]{IEEEtran}
\usepackage[nospace]{cite}
\usepackage{amsmath,amssymb,amsfonts}
\usepackage{graphicx}
\usepackage{textcomp}
\usepackage{xcolor}
\usepackage{fancyhdr}
\usepackage[hyphens]{url}

\usepackage{subfigure}
\usepackage{algorithm}
\usepackage{algpseudocode}
\usepackage{multirow}
\usepackage{makecell}

\usepackage{braket}
\usepackage{authblk}

\newcommand{\linq}{LinQ}
\newcommand{\ltqc}{TILT}
\newcommand{\tilt}{TILT}

\def\BibTeX{{\rm B\kern-.05em{\sc i\kern-.025em b}\kern-.08em
    T\kern-.1667em\lower.7ex\hbox{E}\kern-.125emX}}

\title{TILT: Achieving Higher Fidelity on a Trapped-Ion Linear-Tape Quantum Computing Architecture} 
\author[1,*]{Xin-Chuan Wu\thanks{* Corresponding author: xinchuan@uchicago.edu}}

\author[2]{Dripto M. Debroy}
\author[1]{Yongshan Ding}
\author[1]{Jonathan M. Baker}
\author[5]{\\Yuri Alexeev}
\author[2,3,4]{Kenneth R. Brown}
\author[1]{Frederic T. Chong}

\affil[1]{Department of Computer Science, University of Chicago, Chicago, IL 60637, USA}
\affil[2]{Department of Physics, Duke University, Durham, NC 27708, USA}
\affil[3]{Department of Electrical and Computer Engineering, Duke University, Durham, NC 27708, USA}
\affil[4]{Department of Chemistry, Duke University, Durham, NC 27708, USA}
\affil[5]{Computational Science Division, Argonne National Laboratory, Lemont, IL 60439, USA}

\begin{document}
\maketitle
\pagestyle{plain}


\begin{abstract}
Trapped-ion qubits are a leading technology for practical quantum computing. In this work, we present an architectural analysis of a linear-tape architecture for trapped ions. In order to realize our study, we develop and evaluate mapping and scheduling algorithms for this architecture.

In particular, we introduce TILT, a linear ``Turing-machine-like'' architecture with a multilaser control ``head'', where a linear chain of ions moves back and forth under the laser head.  We find that TILT can substantially reduce communication as compared with comparable-sized Quantum Charge Coupled Device (QCCD) architectures.  We also develop two important scheduling heuristics for TILT.  The first  heuristic  reduces the number of swap operations by matching data traveling in opposite directions into an ``opposing swap'', and also avoids the maximum swap distance across the width of the head, as maximum swap distances make scheduling multiple swaps in one head position difficult.  The second heuristic minimizes ion chain motion by scheduling the tape to the position with the maximal executable operations for every movement. We provide application performance results from our simulation, which suggest that TILT can outperform QCCD in a range of NISQ applications in terms of success rate (up to 4.35x and 1.95x on average). We also discuss using TILT as a building block to extend existing scalable trapped-ion quantum computing proposals. 

\end{abstract}

\section{Introduction}
Quantum computing (QC) aims to solve certain computational problems beyond the capabilities of even the largest classical high-performance computers. By leveraging the quantum mechanical principles of superposition and entanglement, QC algorithms have the potential to revolutionize areas such as machine learning \cite{biamonte2017quantum}, quantum chemistry \cite{peruzzo2014variational, kandala2017hardware}, and cryptography \cite{shor1999polynomial}. Recently, IBM, Rigetti, and Google demonstrated their QC devices up to 72 superconducting transmon qubits \cite{IBM, Google, Rigetti}, while IonQ and Honeywell have recently made significant steps with trapped-ion QC devices \cite{pino2020demonstration, ionq}. Current QC machines in the Noisy Intermediate-Scale Quantum (NISQ) era \cite{preskill2018quantum} are too small for large benchmarks and unable to support quantum error correction (QEC) \cite{bennett1996mixed,gottesman2010introduction}. We can, however, run on the order of hundreds of quantum operations using on the order of hundreds of quantum bits (qubits).

Trapped-ion technologies are among the most promising systems for scalable QC for many reasons. While most current technologies have limited hardware qubit connectivity, two-qubit gates on trapped-ion computers can be executed on arbitrary pairs of qubits \cite{grzesiak2019efficient}. In a trapped-ion quantum computer, the internal states of the ions form the qubit states and quantum gates are implemented through laser-based operations \cite{grzesiak2019efficient, wright2019benchmarking, bruzewicz2019trapped, lin2009large}, where acousto-optic modulators (AOMs) apply lasers with carefully modulated amplitudes and frequencies in order to generate different quantum gates. In many trapped-ion technologies, the ions themselves must be physically moved, for example, shuttled across the linear trap storing all ions. Previous studies have shown that this transportation in a linear array can be done with minimal energy gain and without the loss of qubit coherence \cite{rowe2002transport, huber2008transport,barrett2004deterministic,hensinger2006t,stick2006ion,kaufmann2017scalable,blakestad2009high}. 

\begin{figure}[!t]
\centering
\includegraphics[width=0.4\textwidth,keepaspectratio]{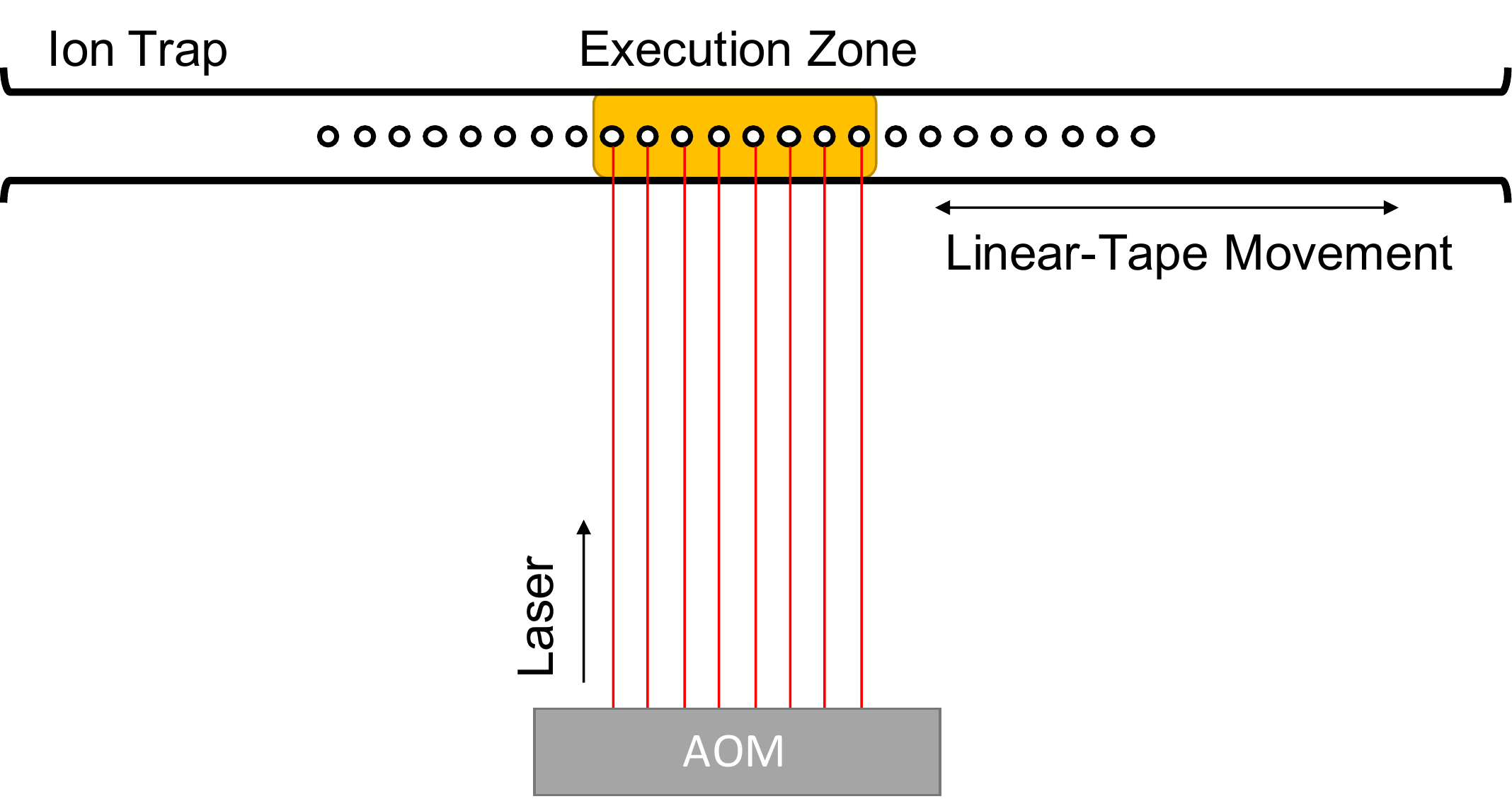}
\caption{A trapped-ion linear-tape quantum computing architecture. Acousto-optic modulators (AOMs) target laser beams for quantum operations, which can be applied to ions in the execution zone. In order to perform gate operations on the other qubits, the entire ion chain is translated until the target qubit is moved into the execution zone.}
\label{fig:iontrap}
\end{figure}

To scale the QC device, we can add ions to the chain, without fear of frequency collision since all ions are identical. While adding ions is relatively simple, controlling all the ions simultaneously is challenging. Recently, a linear-tape architecture was suggested by Monroe \cite{CMonroe2019}, where the device had a fixed set of lasers (often much smaller than the total number of ions in the trap) used to operate all qubits. While the lasers are fixed in place, the ions can be shuttled across the tape to allow them to be operated on. Figure~\ref{fig:iontrap} illustrates this concept, where the \textit{tape} is a string of ions (each representing a qubit) arranged in a single linear trap and the \textit{tape head} is the set of control lasers used to manipulate the qubits. The location of the tape head designates the \textit{execution zone}, or the set of locations where qubits must be located in order for gates to be executed. If a qubit is not located in the execution zone, before applying a gate we must physically move the ions until they are aligned with the tape head. Since the number of ions in the execution zone is limited at one time, the entire ion chain will be moved back and forth during the execution of quantum circuits, similar to a Turing machine \cite{hennie1965one}. While the linear-tape model is inefficient in classical computing, our evaluation shows that a quantum version, with modest parallelism in the tape head, is competitive with contemporary 2D quantum computing designs. The key is that the qubits under the tape head form a completely connected graph and that this powerful communication structure can slide across the entire tape.  Shuttling a small chain of ions has been demonstrated in \cite{fallek2016transport}, and larger tape-like demonstrations are planned \cite{staq2020}.



By sharing laser controls, a trapped-ion linear-tape (\tilt) architecture has easier calibration and  simplified optics, as well as reduced hardware costs. 
In a \ltqc{} system, gates are  implemented only on the qubits within the execution zone, and the entire chain shifts back and forth to operate all qubits. Such a machine does not require the difficult shuttling primitives of a quantum charge-coupled device (QCCD) architecture~\cite{qccd1}, since full chain shuttling is not nearly as difficult as split/merge operations or shuttling over junctions. Moreover, since the ions in the center of a trap are more evenly spaced (which can be accentuated by trap design), such an architecture has fewer issues with individual addressing and laser pointing errors. Consequently, gate fidelities would improve in a \tilt{} architecture, and control optics would be required only for a much smaller region of the chain, resulting in reduced control complexity and cost. 

The construction of trapped-ion quantum computers is heavily influenced by commodity technology, just as classical computing architectures have been. In particular, trapped-ion systems exploit lasers and AOMs used for writing chip masks for commodity silicon chips. The frequency of these commodity lasers determines the ions used (Ytterbium); but more significantly, the size of the AOMs limits the size of the control head to 32 lasers.

\begin{figure}[!t] \centering
\subfigure[Applying a two-qubit gate with swap]
{
\includegraphics[width=0.46\textwidth,keepaspectratio]{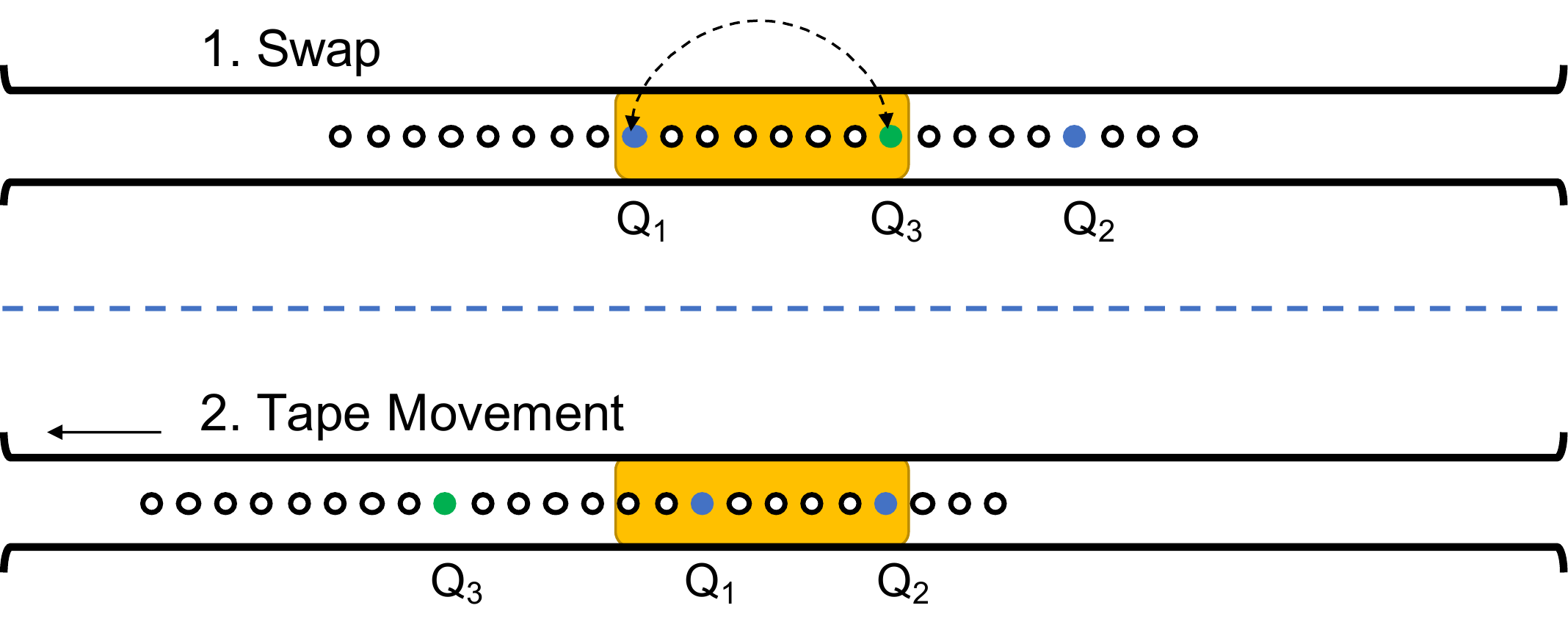}
\label{fig:swap}
}
~\vspace{0.3mm}
\subfigure[Regular swap]
{
\includegraphics[width=0.46\textwidth,keepaspectratio]{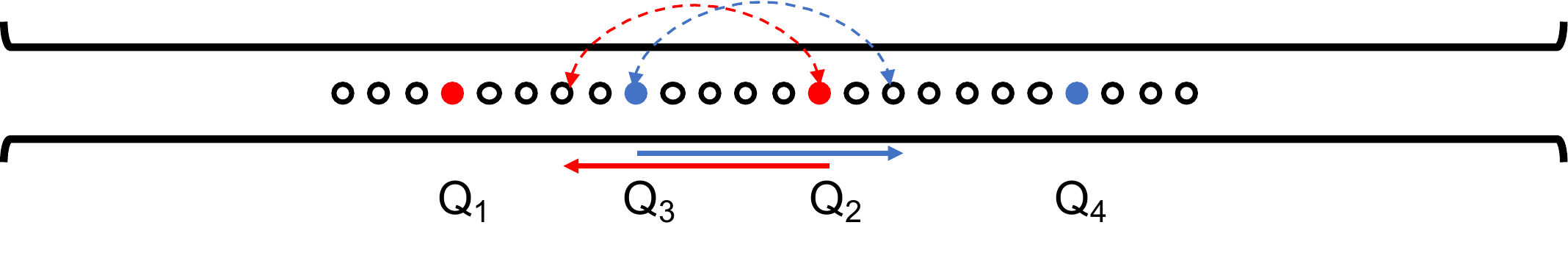}
\label{fig:regular}
}
~\vspace{0.3mm}
\subfigure[{Opposing swap}]
{
\includegraphics[width=0.46\textwidth,keepaspectratio]{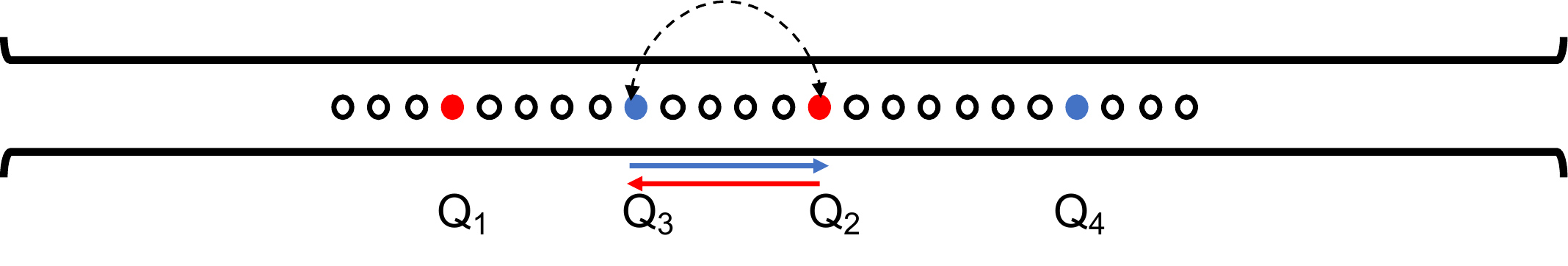}
\label{fig:opposing}
}
\caption{(a) A two-qubit gate is applied to $Q_1$ and $Q_2$ (marked by blue). Since the distance between $Q_1$ and $Q_2$ is longer than the execution zone, a swap between $Q_1$ and $Q_3$ (marked by green) is needed. After the swap, a tape movement brings $Q_1$ and $Q_2$ in the execution zone for the two-qubit gate execution. (b) Suppose a two-qubit gate is applied to \{$Q_1$, $Q_2$\} (marked by red) and another two-qubit gate is applied to \{$Q_3$, $Q_4$\} (marked by blue). Two regular swaps are needed to make the two-qubit gates executable. (c) An opposing swap combines two regular swaps in opposite directions, and this swap makes both two-qubit gates executable. Thus, creating opposing swaps can reduce the total number of swaps.}
\label{fig:swaps}
\end{figure}

In a \ltqc{} system, since the tape head is not covering the entire ion chain, full connectivity is no longer supported among all qubits. For a two-qubit gate, if the distance between the two involved qubits is less than or equal to the tape head size, this two-qubit gate is executable. It may require the whole chain to be moved under the head, but no other gates are required. However, if the distance between the pair of qubits is more than the tape head size, both swap gates and tape movements are necessary, as in Figure~\ref{fig:swap}.

A linear architecture makes communication via qubit swaps more efficient by channeling movement in one path, forcing data to pass in opposite directions; and requiring fewer swap gates involved in the program execution, achieving higher circuit fidelity. Utilizing this architectural feature, we can frequently pair up regular swaps to create \emph{opposing swaps}. As Figure~\ref{fig:opposing} shows, an opposing swap combines two swaps in opposite directions, and hence each opposing swap is equivalent to two regular swaps. This effect, combined with modest tape head sizes, can substantially reduce swaps as compared with 2D architectures.

Regarding the ion chain movement, our scheduling objective is to maximize the overall quantum program success rate. Previous noise analysis studies showed that motional excitations during shuttling are strongest when the acceleration of an ion is the highest \cite{wu2018noise,walther2012controlling}. As a result, the distance traveled, which is generally traversed at a nearly constant velocity, is not particularly relevant to the heating. Instead, there is a chance of heating when the ion begins or ends its motion. When the ion chain starts and stops moving, the movement will trigger the possibility of an increase in thermal motion. This thermal motion can introduce errors during the application of multiqubit gates, as discussed in Section \ref{sec:simulation}. Thus, minimizing the number of movements can improve the overall circuit success rate. 

To support our analysis on \ltqc{}, we have developed \emph{\linq}, a toolflow to compile from high-level quantum programs to a \ltqc{} architecture that minimizes the number of swaps and tape moves and hence improves the expected success rate. We implement a simulator for \ltqc{} to evaluate application metrics by using a suite of NISQ applications. We demonstrate that the \ltqc{} architecture outperforms QCCD architectures in terms of program success rate in a range of NISQ benchmarks. Figure~\ref{fig:linq} shows the LinQ overview.

The main contributions of this work are summarized as follows.

\begin{itemize}
\item To evaluate the feasibility of \tilt{} systems, we present, for the first time, a comprehensive design and evaluation of the \ltqc{} architecture targeting systems with 64 qubits. 

\item To evaluate the performance of \tilt{} architectures, we develop our toolflow, \linq, that provides an optimizing compiler and simulator. We develop swap insertion and shuttling strategies to improve the overall quantum program success rates.

\item To precisely understand the impact of thermal heating, we simulate using a noise model extracted from realistic experimental studies to estimate the application reliability.

\item Our simulations show that \ltqc{} can achieve higher program success rates on a range of NISQ benchmarks compared to QCCD systems (up to 4.35x and 1.95x on average). Our results suggest that \tilt{} provides a viable path toward scalable quantum computers. 
\end{itemize}


\section{Background}\label{sec:background}
In this section, we give a brief overview of quantum computation. We then present the relevant background on trapped-ion systems.

\subsection{Principles of Quantum Computation}
\noindent\textbf{Qubit.} A qubit is a two-level quantum system usually defined by two computational orthonormal basis states $\ket{0}$ and $\ket{1}$. A quantum state can be expressed by any linear combination of $\ket{0}$ and $\ket{1}$: $\ket{\psi} = \alpha\ket{0} + \beta\ket{1}$, where $\alpha$ and $\beta$ are complex amplitudes satisfying $|\alpha|^2 + |\beta|^2 = 1$.

\noindent\textbf{Quantum Gates.} Quantum gates are unitary operations applied on qubits to map from one quantum state to another. Gates are applied to one or multiple qubits simultaneously. Arbitrary single-qubit gates and two-qubit controlled gates are known to be universal \cite{divincenzo1995two}. A quantum circuit (application) consists of a sequence of quantum gates to evolve the quantum state toward the solution state.

\subsection{Trapped-Ion QC Systems}
In a trapped-ion quantum computer, information is stored in the internal states of atomic ions which are trapped in an oscillatory radio-frequency magnetic field. This field constrains the positions of the ions, and almost all current proposals have the ions arranged in a linear crystal, commonly referred to as the \textit{ion chain}. The internal states of the ions can be manipulated by shining lasers on individual ions in order to implement single-qubit gates, and the motion of the chain is used as a bus in order to mediate entangling operations between multiple qubits. The qubit can be defined by either the hyperfine states or Zeeman states of the ion, each of which has benefits beyond the scope of this study~\cite{PhysRevA.97.052301}.

\noindent\textbf{Ion Chain and Qubit Characteristics.} 
The ions in most modern ion traps are spaced approximately $5$ microns apart. We do not consider $T_1$ or $T_2$ times in this study, but for a qubit built off of the hyperfine states of $Yb_{+}^{171}$ ions, the former is on the order of $10^{11}s$ and the latter is generally on the order of seconds, but measurements of up to ten minutes have been achieved~\cite{wang2017single}. The state of the qubits in this chain can be measured by shining a laser on the beam, which causes ions in the $|1\rangle$ state to fluoresce. The ion crystal can then be imaged, with ions in the $|1\rangle$ state appearing as bright spots, and ions in the $|0\rangle$ state appearing as dark gaps~\cite{ozeri2011trapped}.


\noindent\textbf{Laser-Based Operation.}
In a trapped-ion quantum computer, gates are applied via laser pulses aimed at the target ions. Different operations can be implemented by varying the intensity, frequency, phase, or length of these pulses. For example, consider a single qubit gate. A laser is aimed at the ion which causes its state to slowly rotate, in the Bloch sphere, about the desired Pauli axis. The rotation speed can be increased by increasing the intensity of the beam. By carefully monitoring the intensity of the laser and tracking the timing, the desired rotation can be implemented. Two qubit gates are more complex; for trapped-ion quantum computers, the canonical two-qubit gate is the M\o{}lmer-S\o{}rensen gate, which implements an $XX(\theta) = \exp(i\frac{\theta}{2}\mathbf{XX})$ operation by using lasers individually addressed on the desired pair of qubits, along with a global beam that hits all qubits in the chain. In order to create long-range two-qubit entanglement, this gate entangles each qubit with the motional state, applies state-dependent forces, and then unentangles the qubits from the motion. If done perfectly, this process leads to ions whose internal states are entangled without any residual entanglement between the internal states and the motion. In modern experiments, raw single-qubit gates have error rates around $10^{-3}$, although by using composite pulses these rates can be improved significantly~\cite{brown2004arbitrarily, wimperis1994broadband}. Error rates on two-qubit gates often depend on the length of the chain; but in small experiments, numbers as low as $10^{-3}$ have been reported, although most experiments are still on the order of $10^{-2}$~\cite{leung2018robust,PhysRevLett.117.060504,wright2019benchmarking}.


\noindent\textbf{Tape Shuttling.}
In the proposed \ltqc{} architecture, the physical ion chain takes the place of the linear tape. By modulating the DC voltages of the electrodes in the trap, the ion chain can be moved along the axis of the trap. As explained above, lasers are used to process the information in a trapped-ion quantum computer. The idea behind an \ltqc{} architecture is that operations are applied only to ions near the middle of the trap, in the \emph{execution zone}, and the chain is shifted back and forth in order to make long-range interactions occur. This strategy is counter to another architectural proposal, the quantum charge-coupled device~\cite{qccd1}. A QCCD architecture comprises many smaller trapping zones within a single chip, and ions are shuttled around individually. Recently, Honeywell built the first QCCD system with four qubits~\cite{pino2020demonstration}. While such an architecture allows for more parallelization and flexibility, it has high costs in terms of shuttling complexity since it requires expensive split/merge and junction crossing maneuvers, which lead to more thermal energy entering the system. 

Thermal energy is stored in the motion of the chain, and it is also a significant factor in \ltqc{} architectures. The M\o{}lmer-S\o{}rensen gate mentioned in the preceding section is designed to be insensitive to the motional energy of the chain; however, this is true only for a perfectly applied gate. In reality, there is a small contribution to the infidelity of the gate, which is caused by residual entanglement between the internal state of the ions and the motional state, which was used to mediate the interaction. This error is due to the imperfect closure of a loop in phase space (the space of the positions and momenta of the particle). The error caused by this imperfect closure scales as $\exp(2n + 1)$, where $n$ is the number of motional quanta in the chain. As a result, as a chain heats up, it becomes more sensitive to imperfections in the application of the gate.


\section{\tilt{} Architecture}\label{sec:architecture}

\tilt{} architectures have a linear chain of ions trapped within the oscillating potential which shuttles as one large block. We call this linear chain a \textit{tape}. Most shuttling studies focus on split and merge operations or shuttling single qubits through junctions. A primary benefit of the linear tape architecture is that these difficult maneuvers are unnecessary. 

\subsection{Feasibility}
Prior works~\cite{wu2018noise,walther2012controlling} show that shuttling distance of ions is irrelevant to the heating since ions are shuttled at a nearly constant velocity. However, the start and stop of the ion shuttling will add thermal heating to the ion chain. Based on previous noise analysis studies~\cite{wu2018noise}, we derive a heating model in which every time the chain is shuttled its motional energy increases by some fixed amount $k$. This quantity is affected by how free the chain is to oscillate in its center of mass mode after shuttling. As a result, we scale this value by $k\sim\sqrt{n}$ when moving to a larger chain, where $n$ is the number of ions of the chain. Whenever a two-qubit gate is enacted, its fidelity is dependent on the total motional energy of the chain. High motional energy leads to larger gate errors, so shuttles have a negative impact on the overall success rate of the future gates. As a result, compiling techniques minimizing the number of tape moves are favored.

All components of this architecture have been developed to some extent, however ion trap quantum computing experiments have yet to be attempted at the point where our techniques are necessary. Chains as large as 79 ions have been demonstrated by ionQ~\cite{ionq}, however they didn't individually address all ions. Using a \tilt{} architecture would make this addressability concern much simpler. Shuttling techniques like the ones needed for our work have already been demonstrated for small segments of crystals, and would be scaled up to larger chains without an increase in complexity~\cite{pino2020demonstration, walther2012controlling, PhysRevA.99.022330}. The benefit of the \tilt{} architecture is that it does not require any pieces that have not been demonstrated already at the level of complexity necessary for the architecture, as opposed to QCCD architectures which will need more complex junction shuttles and trap designs as they scale up from their small demonstrations to larger quantum devices.

\subsection{Gate Selection}
There are multiple gate implementations for trapped-ion systems, including Frequency Modulated (FM), Phase Modulated (PM), and Amplitude Modulated (AM) gates~\cite{murali2020architecting,wright2019benchmarking,blumel2019power,ball2020software}. The gate time for FM gates is proportional to the total number of ions in a chain~\cite{leung2018robust,leung2018entangling}. However, the operation time of PM and AM gates is only proportional to the distance between two involved ions~\cite{wu2018noise,trout2018simulating,milne2020phase}. Since \tilt{} has a long chain with a smaller region for operations, using FM gates would not take advantage of our systems structure, leading to extra gate error. Therefore, PM or AM gates are proper gate implementations for \tilt{}. In our study, we consider amplitude modulated (AM) gates for our two-qubit gate implementation.

\subsection{Compared with QCCD}\label{sec:compare}

\begin{figure}[!t] \centering
\subfigure[Qubits communication in QCCD.]
{
\includegraphics[width=0.46\textwidth,keepaspectratio]{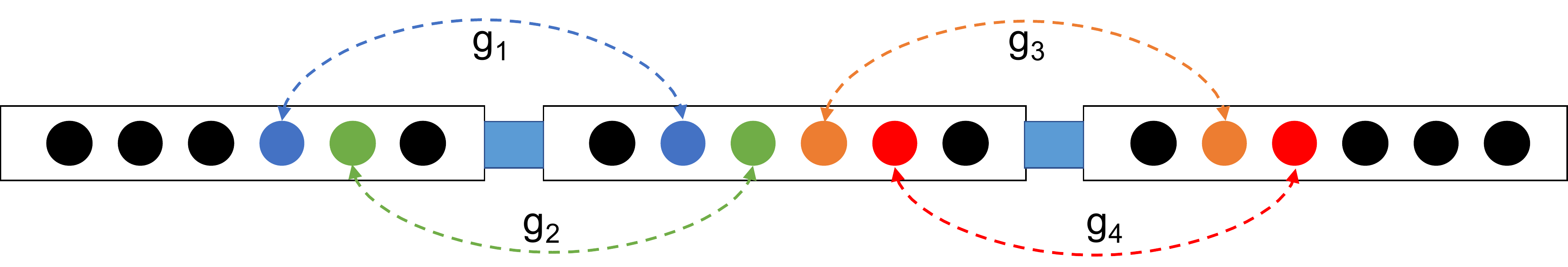}
\label{fig:qccd_compare_1}
}
~\vspace{0.3mm}
\subfigure[Qubits communication in \tilt{}.]
{
\includegraphics[width=0.46\textwidth,keepaspectratio]{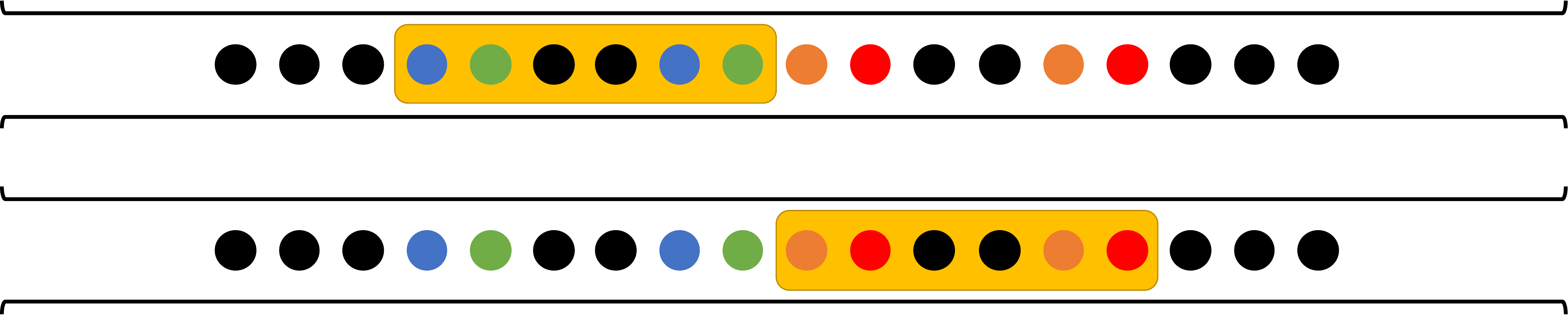}
\label{fig:qccd_compare_2}
}
\caption{ Compare \tilt{} with QCCD. g1, g2, g3, and g4 are two-qubit gates on  four pairs of qubits (marked by blue, green, orange, and red). (a) For QCCD, when a qubit communicates with another qubit in the different trap, QCCD requires the following operations: i) Swap the qubit to the end of the current trap. ii) Split from the current chain. iii) Shuttle to the target trap. iv) Merge to the target chain. v) Interact with the target qubit. To execute the gates g1, g2, g3, and g4, QCCD needs to repeat the above operations four times. (b) For \tilt{}, performing a tape movement would allow the operations of the desired two-qubit gates.}
\label{fig:qccd_compare}
\end{figure}

Another proposed model to scale trapped-ion QC systems is the quantum charge coupled device (QCCD), which requires complex junction shuttles and trap designs as they scale up~\cite{qccd1}. Additionally, QCCD requires ion chain split/merge operations during the process of a computation~\cite{murali2020architecting} for communication with qubits in other traps. 

Figure~\ref{fig:qccd_compare} shows the operations required for a short-distance (within the laser head size) communication pattern. QCCD requires a series of operations to perform all two-qubit interactions. However, \tilt{} only needs a single tape movement to achieve the operations. As a result, \tilt{} is expected to have higher success rates. Furthermore, if there are multiple short-distance communications, a QCCD system would require multiple split/merge/shuttle operations, but it is possible for \tilt{} to achieve these operations with only one tape movement through a well-designed tape move scheduling. In this case, \tilt{} will outperform QCCD architectures. For long-distance communication, QCCD requires almost the same operations for short-distance communication if there is a shuttling pathway, but for \tilt{}, the system might require multiple swaps and tape shuttles. Thus, QCCD can perform better than \tilt{} in dealing with a long-distance communication.

In the current state of QC, most quantum applications heavily rely on short-distance communication patterns, such as the Variational Quantum Eigensolver (VQE)~\cite{kandala2017hardware}, Quantum Approximate Optimization Algorithm (QAOA)~\cite{farhi2014quantum}, Quantum Error Correction (QEC) like the surface code~\cite{fowler2012surface}, Quantum Generative Adversarial Networks (QGAN) \cite{dallaire2018quantum, lloyd2018quantum,situ2018quantum,hu2019quantum,zeng2019learning}, and the Ising model solver \cite{barends2016digitized}. Therefore, \tilt{} is a promising design to achieve higher fidelity on these applications with a trapped-ion quantum computing architecture.

\subsection{\tilt{} Challenges}
Under the device limitations of \tilt{}, there are two major challenges to performing quantum circuits. Since the qubits are not fully connected, two-qubit gates on pairs of qubits whose distance is greater than the head size will need swap gates in order to make two qubits close enough to become executable. Consequently, we need qubit mapping and swap insertion techniques which minimize the number of swaps in order to achieve high quantum program success rates. Second, since only a part of the ion chain can be moved into the execution zone and since the tape movement will cause a heating error, minimizing the number of tape moves is essential to success.

\section{LinQ for \tilt{} Architecture}\label{sec:linq}

To evaluate the \tilt{} architecture, we develop our toolflow, LinQ, including the compiler and simulator. LinQ allows us to compile a quantum program written in a high-level programming language to \tilt{}-architecture-level instructions. LinQ takes the device specification as input, including the number of qubits and the tape head size, and returns a compiled circuit, optimizing for program success rate. Since the \tilt{} machine is not yet realized, we use simulation to predict the performance of this novel architecture. We summarize the notations used in this paper in Table~\ref{tab:notation}.

\begin{table}[h!]
\centering
\caption{Notations used in this paper.}
\small
\begin{tabular}{ll}
\hline\hline
Notation & Definition\\
\hline
$n$ & The number of ions of a chain.\\
\hline
$g$ & A two-qubit gate. \\
\hline
$g.q_1, g.q_2$ & A pair of qubits that $g$ is applied to.\\
\hline
$d_g$ & The distance between $g.q_1$ and $g.q_2$\\
\hline
$M$& A mapping from logical to physical\\
\hline
$M_{q_i,q_j}$ & A qubit mapping after swapping $(q_i, q_j)$\\
\hline
$G$ & A set of two-qubit gates\\
\hline
$D(g,M_{q_i,q_j})$ & $d_g$ under the mapping $M_{q_i,q_j}$\\
\hline
$\Gamma$ & Background heating rate of the trap\\
\hline
$\tau$ & Gate time\\
\hline
$k$ & Amount of heating added during each shuttle\\
\hline
$\epsilon$ & Gate error due to residual entanglement\\
\hline\hline
\end{tabular}
\label{tab:notation}
\end{table}

\subsection{LinQ Overview}\label{sec:linq_overview}
\begin{figure}[!t]
\centering
\includegraphics[width=0.38\textwidth,keepaspectratio]{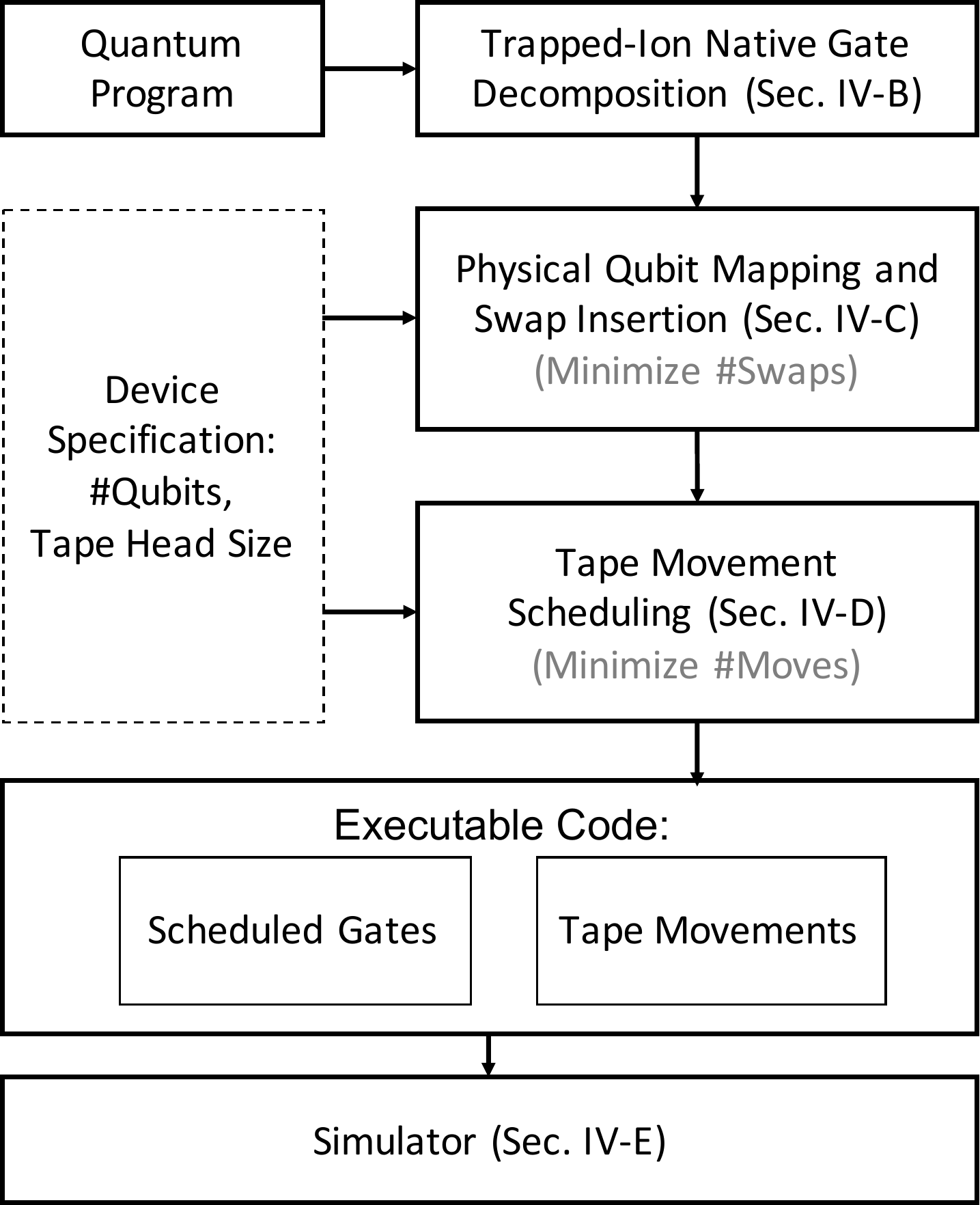}
\caption{LinQ overview. Taking a quantum program and device specification as input, LinQ generates the executable codes, including a sequence of scheduled gates and tape movements, and then runs the simulation to output the success rate and run-time.}
\label{fig:linq}
\end{figure}

Figure~\ref{fig:linq} shows an overview of our toolflow. LinQ contains three core compiler modules: trapped-ion native gate decomposition, qubit mapping and swap insertion, and tape movement scheduling. Since our target device size for \tilt{} is more than 60 qubits, using one procedure for optimizing qubit mapping, swap insertion, and tape movement scheduling will have a long compilation time. Therefore, we separate the optimization problem into two steps and address them independently. However, when we perform swap insertion, we should take tape movement scheduling into consideration, because the swap patterns will affect the number of tape movements. In Section~\ref{sec:eval_arch} we will show that this two-step optimization can generate success rates similar to those of ideal trapped-ion systems.

The inputs for LinQ consist of a high-level quantum program and the device specification, including the number of qubits in the tape and the tape head size. First, LinQ decomposes the quantum program to trapped-ion native gates. Next, LinQ maps the logical qubits to physical qubits and resolves all long-distance two-qubit gates by inserting swap gates. Then, LinQ performs our tape movement scheduling algorithm. The output is the executable circuit consisting of a sequence of scheduled gates and tape movements. 

To evaluate the performance, such as success rates and the execution time, we pass the sequence of scheduled gates and tape movements to our simulator, which uses noise models extracted from realistic experimental studies.

\subsection{Native Gate Decomposition}\label{sec:linq_decomposition}
In order to generate executable code, gates from high-level quantum programming languages must be decomposed into the device-dependent native gates. LinQ decomposes the universal gates, such as CNOTs, CZs, and other operations into the \tilt{} native gate set \cite{maslov2017basic}. For example, we decompose a CNOT $q_1,q_2$ gate into a sequence of rotations and XX operations: $R_y(\pi/2) \ q_1;$ $XX(\pi/4)\ q_1, q_2;$ $R_x(-\pi/2)\ q_1;$ $R_x(-\pi/2)\ q_2;$ $R_y(-\pi/2)\ q_1$. 

\subsection{Qubit Mapping and Swap Insertion}\label{sec:linq_swap}

Unlike fully connected trapped-ion systems, \tilt systems cannot operate long-distance two-qubit gates directly if the distance between the pair of qubits is longer than the tape head size. These long-distance two-qubit gates become unexecutable gates and to resolve them, we must insert swap gates. For a two-qubit gate $g$ applied on $q_1$ and $q_2$, we use $d_g$ to denote the distance between $q_1$ and $q_2$. The swap insertion problem is to find a sequence of intermediate qubits, for example, $\{(q_1, q_i), (q_i, q_j), \dotsc, (q_k, q_2)\}$, such that any pair $(q_i, q_j)$ in the sequence has distance smaller than the tape head size.

Several approaches of swap insertion have been proposed. One common approach is to formulate the problems in a mathematical form, such as integer linear programming, and then utilize software solvers to find the optimal solutions~\cite{bahreini2015minlp,wille2014optimal,lye2015determining,maslov2008quantum,wu2019ilp}. This method is guaranteed to provide an optimal solution. However, since the time-to-solution grows exponentially with the number of qubits and the circuit depth (or the number gates), scaling this approach to large NISQ programs is infeasible. In our study, we focus on 60+ qubits applications. Hence, we need a practical solution for the swap insertion problem.

\noindent\textbf{Baseline Approach.} A straightforward idea is to allow swap with the tape head size distance to ensure a minimal number of swaps required for a two-qubit gate. We firstly establish our baseline implementation by using IBM Qiskit \emph{StochasticSwap}~\cite{cross2018ibm}, which is commonly used for swap insertion and it is also used in the highest optimization level of Qiskit circuit compilation pass. 

However, two problems arise. First, using the tape head size as the swap distance will force the tape to move exactly once for every swap. If $d_g$ is less than the tape head size, the tape movement scheduler will have multiple choices to schedule the tape position for gate $g$; If $d_g$ is equal to the tape head size, the scheduler must move the tape to the exact position for the two-qubit gate, and hence this will increase the number of tape movements in many cases. Second, this swap insertion method does not consider opposing swaps. An opposing swap (Figure~\ref{fig:opposing}) combines two regular swaps in opposite directions. As a result, creating opposing swaps frequently can reduce the total number of swaps. This reduced swap count results in greater circuit success rates.

A good qubit mapping algorithm can also reduce the number of swap gates. Previous studies have proposed two types of solutions for qubit mapping and swap insertion for 1D and 2D architectures. One is to formulate the problem as an equivalent mathematical problem and then use a solver to find the optimal mapping to the problem \cite{maslov2008quantum,chakrabarti2011linear}. This approach cannot scale to solve large problem for similar reasons to the swap insertion problem. The other solution is to use a heuristic search to obtain a mapping \cite{nishio2019extracting,li2019tackling, itoko2020optimization}. Since our target size is 64 qubits, we use a heuristic search to address this problem.

\begin{figure}[!t]
\centering
\includegraphics[width=0.42\textwidth,keepaspectratio]{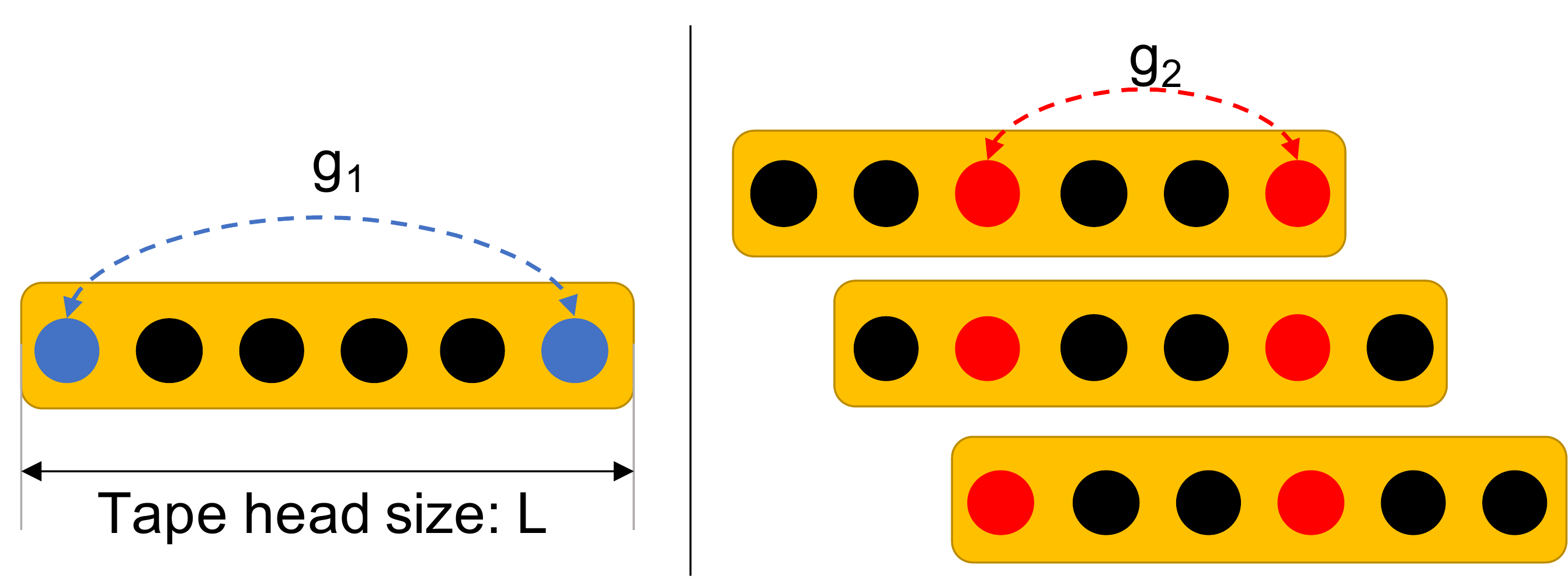}
\caption{Considering two-qubit gates $g_1$ and $g_2$ on a system with the tape head size of $L$. Since $d_{g_1}$ is $L - 1$, only one tape head position is allowed for $g_1$ to be executed. As for $g_2$, there are three valid tape head positions because $d_{g_2}$ is $L - 3$.}
\label{fig:max_swap_length}
\end{figure}

\noindent\textbf{LinQ Approach.} We adopt the existing heuristic algorithms for initial qubit mapping \cite{li2019tackling, itoko2020optimization}. After we have an initial qubit mapping $M$, we improve the swap insertion. For resolving an unexecutable two-qubit gate $g$ on $q_1$ and $q_2$, we insert swaps by applying our heuristic algorithm, shown in Algorithm~\ref{algo:swap}. The idea is to create opposing swaps to reduce the overall swap count. For each two-qubit gate applied on $q_1$ and $q_2$, for all $q_i$ between $q_1$ and $q_2$, if the distance between $(q_i, q_1)$ or $(q_i,q_2)$ is less than the maximal swap length ($MaxSwapLen$), then we put the swap candidate into a list $S$. Next, we compute the score for each swap candidate and select the swap with the minimal score to update the mapping and insert the swap to the circuit. If the tape head size is $L$, $MaxSwapLen$ can be $L - 1$. For a swap gate $g$, if $d_g$ is equal to $L - 1$, the tape must be moved to the exact position for the gate execution, as shown in Figure~\ref{fig:max_swap_length}. As a result, the circuit might require more tape movements. To mitigate this problem, we restrict $MaxSwapLen$ to a certain number less than $L - 1$ to create the flexibility for our tape movement scheduling. As long as the swap gate count is not increased significantly, by shorting the $MaxSwapLen$, we can increase the available head positions to potentially operate more gates in one tape movement. The best choice of the $MaxSwapLen$ depends on the structure of the target application. An ideal $MaxSwapLen$ will limit the overall $d_g$ but will not increase the total number of swap gates. We can repeatedly run the procedure with different parameters to select the best $MaxSwapLen$.

\begin{algorithm}[t]
 \caption{Swap Insertion}\label{algo:swap}
 \begin{algorithmic}[1]
 \State $S \gets \phi$
  \While { $g$ is unexecutable}
    \State $C \gets \phi$
    \For {$q_i$ from $g.q_1$ to $g.q_2$}
      \If {$dist(q_i, g.q_1) < MaxSwapLen$}
        \State $C \gets C \cup \{(q_i, g.q_1)\}$
      \EndIf
      \If {$dist(q_i, g.q_2) < MaxSwapLen$}
        \State $C \gets C \cup \{(q_i, g.q_2)\}$
      \EndIf
    \EndFor
    \For {$(q_i,q_j)$ in $C$}
      \State Compute $Scores(M_{q_i, q_j})$
    \EndFor
    \State Find the swap with minimal score
    \State $S.append(Swap(q_i, q_j))$
    \State $M \gets M_{q_i, q_j}$
  \EndWhile
  \State Insert the swap gate sequence $S$ to the circuit
 \end{algorithmic}
\end{algorithm}

The heuristic score function sums $d_g$ for each two-qubit gate $g$ under a mapping $M$. If a swap pair ($q_i, q_j$) leads to the minimal sum of all $d_g$, this swap pair will be selected. The swap score of the qubit pair $(q_i,q_j)$ is defined as follows:

\begin{equation}
 Score(M_{q_i, q_j}) = \sum_{g\in G}D(g,M_{q_i, q_j})\times \alpha^{\Delta(g)},
\end{equation}
where $G$ is the set of remaining two-qubit gates, $M_{q_i, q_j}$ is the qubit mapping after swapping $q_i$ and $q_j$, $D(g,M_{q_i, q_j})$ denotes $d_g$ under the mapping $M_{q_i, q_j}$, $\alpha$ is a parameter ($0 < \alpha < 1$) used to prioritize two qubit gates, and $\Delta(g)$ is the circuit depth distance between the gate $g$ and the current resolved gate. We use the sum of $d_g$ to consider the impact for the future swaps. The optimal swap can impact the swaps in the future to reduce the overall qubit distance between the pair of qubits. Thus, we choose the swap candidates have the minimal scores.  

The computational complexity of our heuristic algorithm can be estimated by the worst case, where the two qubits are separated in two ends and the swap insertion is performed repeatedly until two qubits are close enough to make the two-qubit gate become executable. The complexity is $O(NG)$ for a two-qubit gate, where $N$ is the total number of qubits in a tape and $G$ is the number of remaining two-qubit gates.

\subsection{Tape Movement Scheduling}\label{sec:linq_move}
In \tilt{} architectures, since the tape head is shared among all qubits, the tape will be moved back and forth frequently to execute a quantum application. Such tape movement/shuttling can introduce noise into the systems. Aware of issues like this, an effective tape movement scheduler can increase circuit fidelity by mitigating shuttling noise. In this work, we model the fidelity of the operations in order to improve the success rate. Our noise model considers the impact of thermal heating due to shuttling~\cite{wu2018noise}. Since every tape move introduces thermal heating to the \tilt{} system, this introduced heating leads to a lower gate fidelity. Hence, The design of our scheduling algorithm is to perform as many operations as possible before a tape movement is performed. Algorithm~\ref{algo:move} shows the pseudocode of our scheduling algorithm. The \emph{ExecutableGates} follows the gate dependency order. We compute the scores for each tape head position and select the position with the maximal score to schedule; we then repeat the procedure until all gates are scheduled and executed. 

\begin{algorithm}[t]
 \caption{Tape Movement Scheduling} \label{algo:move}
 \begin{algorithmic}[1]
 \State $T \gets \phi$
 \State $E \gets \phi$
 \State $R \gets G$
 \While {$R$ is not empty}
  \For {each tape head position $p_i$}
    \State $F_i \gets \{ExecutableGates\}$
    \State Compute $Score(p_i)$
  \EndFor
  \State Find the tape head position $p$ with the maximal score;
  \State $T.append(p)$
  \State $E.append(F_p)$
  \State $R \gets R - F_p$
 \EndWhile
 \State Output the tape movement scheduling $T$ and gate execution sequence $E$
 \end{algorithmic}
\end{algorithm}

The objective of the heuristic cost function is to indicate the number of executable gates for a tape head position. As a result, the number of executable gates is the score for each tape head position. The general form of our heuristic cost function for a tape head position $p$ is shown as follows:
\begin{equation}
 Score(p) = n_{p},
\end{equation}
where $n_{p}$ is the number of executable gates at the position $p$.

The time complexity of our heuristic scheduling algorithm is $O((N-L)DL)$, where $N$ is the number of qubits in a trap, $L$ is the tape head size, and $D$ is the circuit depth. 

\subsection{\tilt{} Simulation}\label{sec:simulation}
To understand the impact of shuttling noise on \tilt{} architectures, we perform noisy circuit simulation using realistic noise models. In our simulations we consider amplitude modulated (AM) gates for our two-qubit gate implementation~\cite{trout2018simulating}. These gates have a gate time of 
\begin{equation}\label{eq:gate_time}
\tau(d) = 38\times d + 10
\end{equation}

where $d$ is the distance between the two ions, in units of ion spacings, and the time is given in microseconds. This same model was studied in \cite{murali2020architecting}.

In an ion-trap quantum computer, single-qubit gate fidelities are independent of the thermal energy in the trap. Additionally, a perfectly applied M\o{}lmer-S\o{}rensen gate is independent of the thermal energy of the chain. As the vibrations increase, however, the gate becomes more sensitive to any imperfections in the laser pulses, which we quantify through the following model of gate fidelities after $m$ moves have occurred, inspired by the work done in \cite{wu2018noise}:
\begin{equation}\label{eq:noise}
  \begin{split}
    F_{m} &= 1 - \Gamma\tau + (1 - (1 + \epsilon)^{2mk + 1})\\
  \end{split}
\end{equation}
where $\Gamma$ is the background heating rate of the trap, $\tau$ is the gate time defined in microseconds as a function of number of intra-ion spacings between the involved qubits~\cite{murali2020architecting}, $k$ is the average amount of heating added to the chain during each shuttle, as discussed in Section~\ref{sec:architecture}; and $\epsilon$ is the error in each two-qubit gate due to residual entanglement with the motion. A gate can be thought of as a loop in phase space, where a perfect gate is a closed loop. An imperfect gate, due to a bad clock, rounding error in the pulse compilation, or laser intensity instability, can lead to this loop not perfectly closing. As the motional mode heats up, the gates become more sensitive to this kind of noise, as explained in~\cite{wu2018noise}. An error model very similar to this is used in \cite{murali2020architecting}. We do not use a linear approximation for the exponential in \cite{wu2018noise} to more accurately model higher motional mode excitations in our device.

In Honeywell's 4-qubit (8-ion) device~\cite{pino2020demonstration}, they report the average heating contribution of their shuttling operations as 2 quanta. This includes primitives such as split/merge and swap, as well as the simpler linear shuttles that the \tilt{} architecture requires. Since split/merge operations and swap operations require complex and precise electrode voltages, these operations have significantly higher contributions to the heating rate than ion chain shuttling. As a result, we can fairly assume that the heating rate due to linear shuttles is lower than this number.

To scale this up to our experiments, we note that the primary consideration for heating is the softening of the common (center of mass) mode as the number of ions in the shuttled chain increases. The stopping force is still caused by the electrodes affecting the end of the chains, and is consequently constant~\cite{walther2012controlling}. However the mass of the chain increases, and as a result the noise in the system scales like a simple harmonic oscillator, with the frequency scaling by $\sqrt{n/F} \sim \sqrt{n}$. 

\section{Experimental Setup}\label{sec:setup}

\subsection{Benchmarks}

\begin{table}[h!]
\small
\centering
\caption{List of benchmarks.}
\begin{tabular}{lccc}
\hline\hline
Application  & Qubits &  2Q Gates  & Communication\\
\hline
ADDER&	64 &	545	&	 Short-distance gates\\
BV&	    64 &	64	 &   Long-distance gates\\
QAOA&   64  &	1260	 &  Nearest-neighbor gates\\
RCS&  	64 	&	560	 &   Nearest-neighbor gates\\
QFT&	64 	&	4032	 &  Long-distance gates\\
SQRT&	78 	&	1028	 &  Long-distance gates\\
\hline\hline\\
\end{tabular}
\label{tab:apps}
\end{table}

To evaluate \tilt{} architectures against real applications, we choose multiple important quantum applications as our benchmarks (Table~\ref{tab:apps}). The benchmarks are selected to have different program characteristics in order to show \tilt{} properties for arbitrary quantum circuits. According to communication patterns, there are three categories of applications. One consists of the applications with long-distance two-qubit gates, which require swap gates to become executable. The second category is for applications with short-distance two-qubit gates and hence do not necessarily need swap gates on TILT. The last category is nearest-neighbor two-qubit gates when the topology is in a 2D grid structure. This category will have short-distance two-qubit communication pattern when mapping on a linear topology.

The adder benchmark is based on the Cucarro adder \cite{cuccaro2004new}. Adders are important kernel functions in many quantum algorithms. Bernstein-Vazirani (BV) is a NISQ application commonly used to benchmark devices \cite{bernstein1997quantum,wright2019benchmarking,debnath2016demonstration}. Quantum Approximate Optimization Algorithm (QAOA)~\cite{farhi2014quantum} benchmark in our evaluation is a hardware-efficiency ansatz~\cite{moll2018quantum} for MaxCut problem. QAOA is a hybrid quantum-classical variational algorithm, and it is one of the most important quantum algorithms in the NISQ era. 
Random Circuit Sampling (RCS) has been proposed by Google to show quantum supremacy \cite{boixo2018characterizing,markov2018quantum,arute2019quantum}.
Quantum Fourier transform (QFT) \cite{javadiabhari2015scaffcc} is an important function in many quantum algorithms (Shor's algorithm \cite{shor1999polynomial}, phase estimation algorithm \cite{cleve1998quantum}, and the algorithm for hidden subgroup problem \cite{jozsa2001quantum}).
This application uses Grover's search algorithm to find the square root number~\cite{javadiabhari2015scaffcc}. Grover's search algorithm is for database search, and it can achieve significant speedups compared with classical search algorithms \cite{grover1996fast, javadiabhari2015scaffcc}.

\subsection{Simulation Parameters}
We evaluate architectures with 60+ qubits and consider tape head sizes of 16 and 32. 
Using the noise model shown in Section~\ref{sec:simulation}, we estimate the fidelity of each gate after $m$ tape moves. 

All experiments are performed on a Ubuntu 16.04 system (Linux kernel 4.4-0-141-generic) with Intel Xeon Silver 4110 32-core CPU at 2.1 GHz and 128 GB of physical memory.
\section{Evaluation}\label{sec:eval}
In our evaluation, we first show the impact of swap insertion. We then compare \tilt{} systems to QCCD systems, and present the compilation and estimated execution times of each application.

\subsection{LinQ Swap Insertion Performance}
In this section, we show the importance of swap insertion choices. We use only those applications with long-distance two-qubit gates (BV, QFT and SQRT) because no swap gates are needed for the other set of applications. We demonstrate the swap impact from two perspectives. First, we compare our LinQ swap insertion heuristic algorithm with a baseline method, which applies Qiskit StochasticSwap compilation tools to swap gate insertion for each unexecutable two-qubit gate. Second, we apply our swap insertion heuristic algorithm with a restricted swap distance to improve the tape movement scheduling. The results with our algorithm show increased circuit success rates.

\begin{figure}[!t] \centering
\includegraphics[width=0.2\textwidth,keepaspectratio]{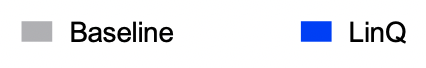}
\\
\subfigure[Opposing swap ratio (Higher is better.)]
{
\includegraphics[width=0.23\textwidth,keepaspectratio]{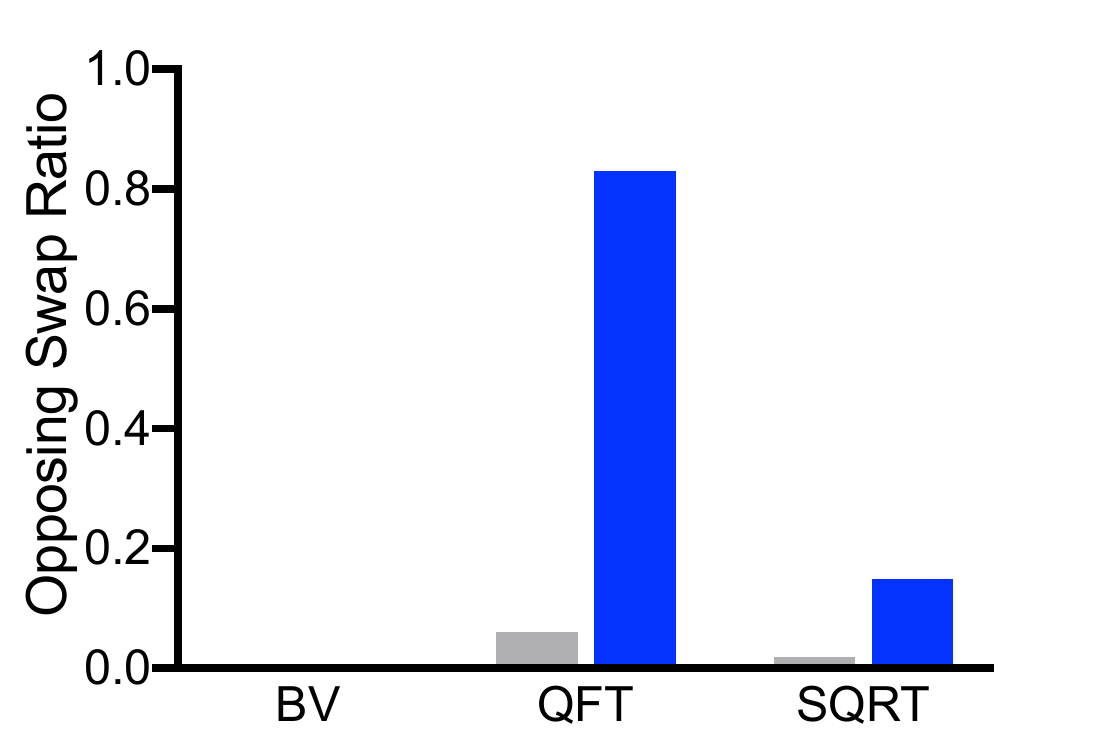}
\label{fig:bo}
}
\hspace{-2mm}
\subfigure[\#Swap (Lower is better.)]
{
\includegraphics[width=0.23\textwidth,keepaspectratio]{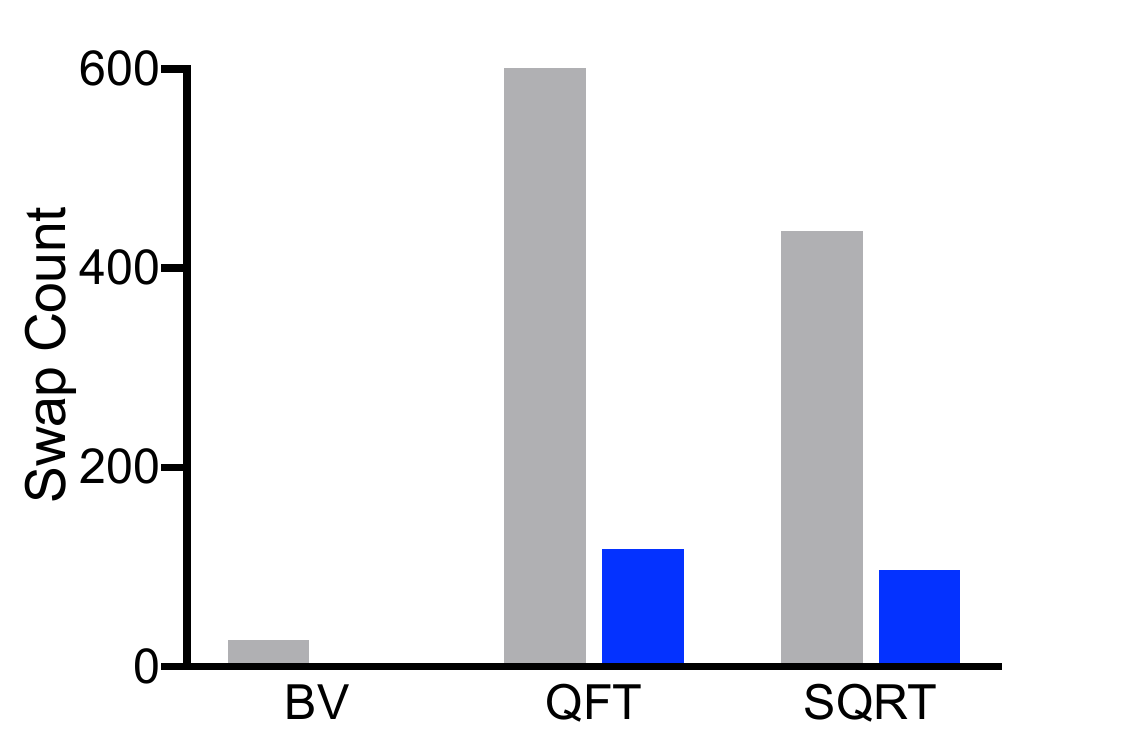}
\label{fig:bs}
}
\hspace{-2mm}
\subfigure[\#Move (Lower is better.)]
{
\includegraphics[width=0.23\textwidth,keepaspectratio]{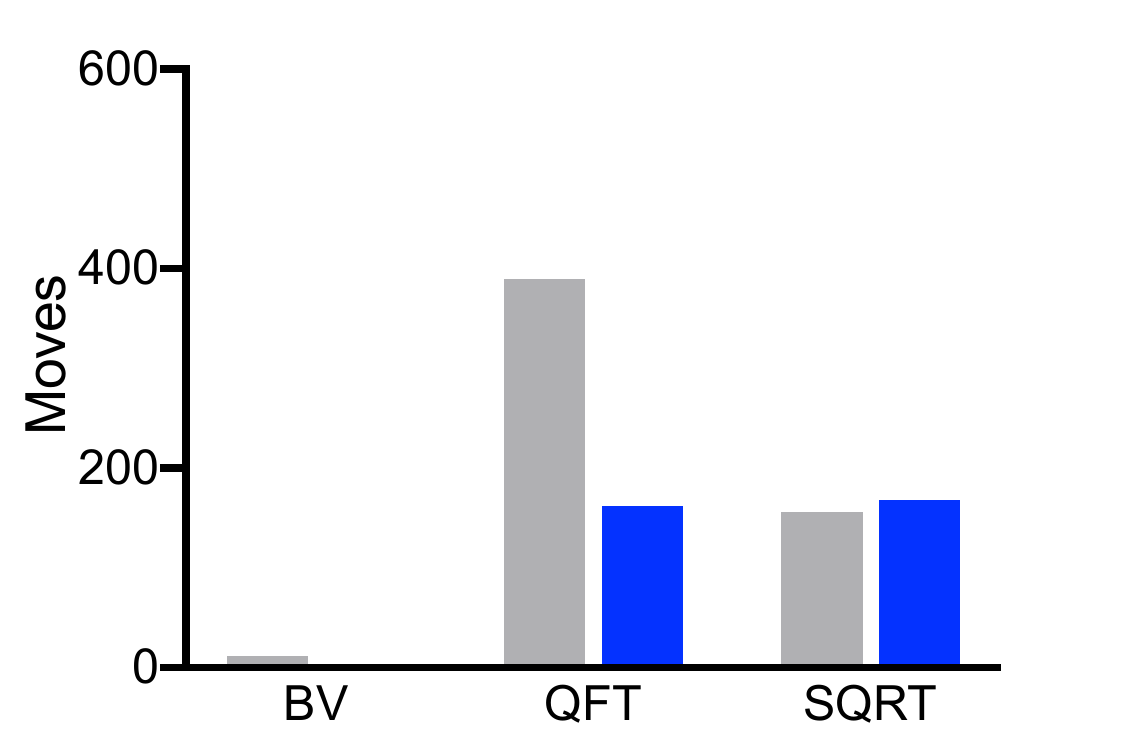}
\label{fig:bm}
}
\hspace{-2mm}
\subfigure[Success rate (BV)]
{
\includegraphics[width=0.23\textwidth,keepaspectratio]{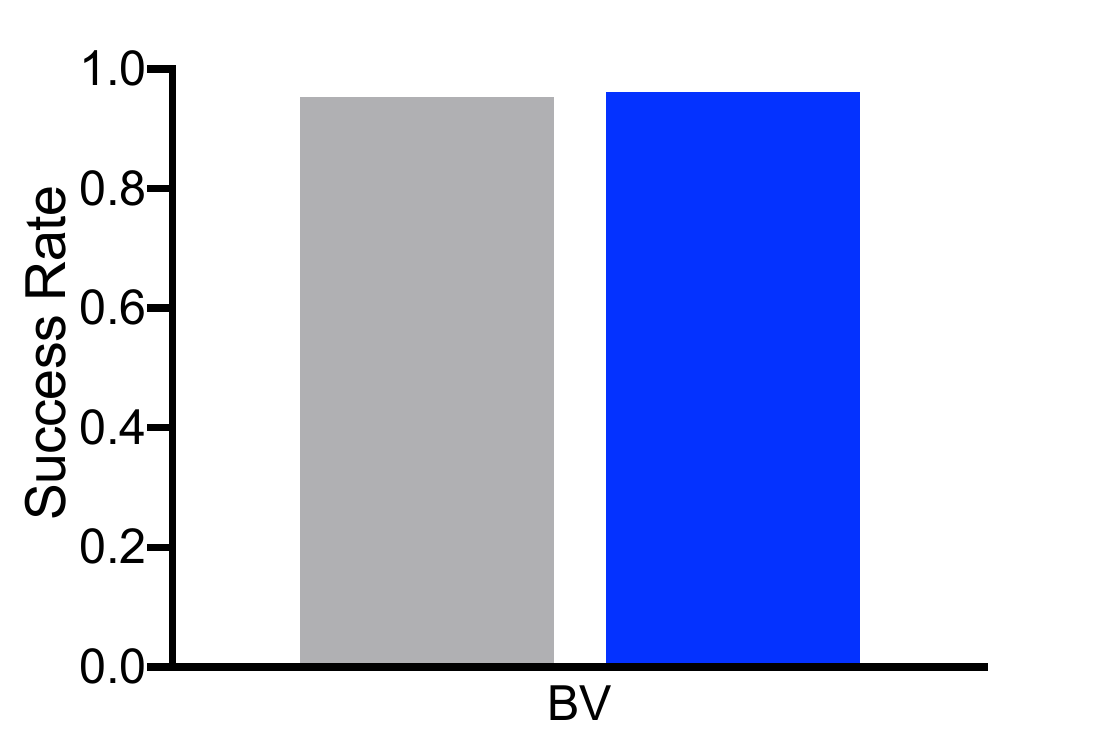}
\label{fig:bsr_bv}
}
\hspace{-2mm}
\subfigure[Success rate (QFT)]
{
\includegraphics[width=0.23\textwidth,keepaspectratio]{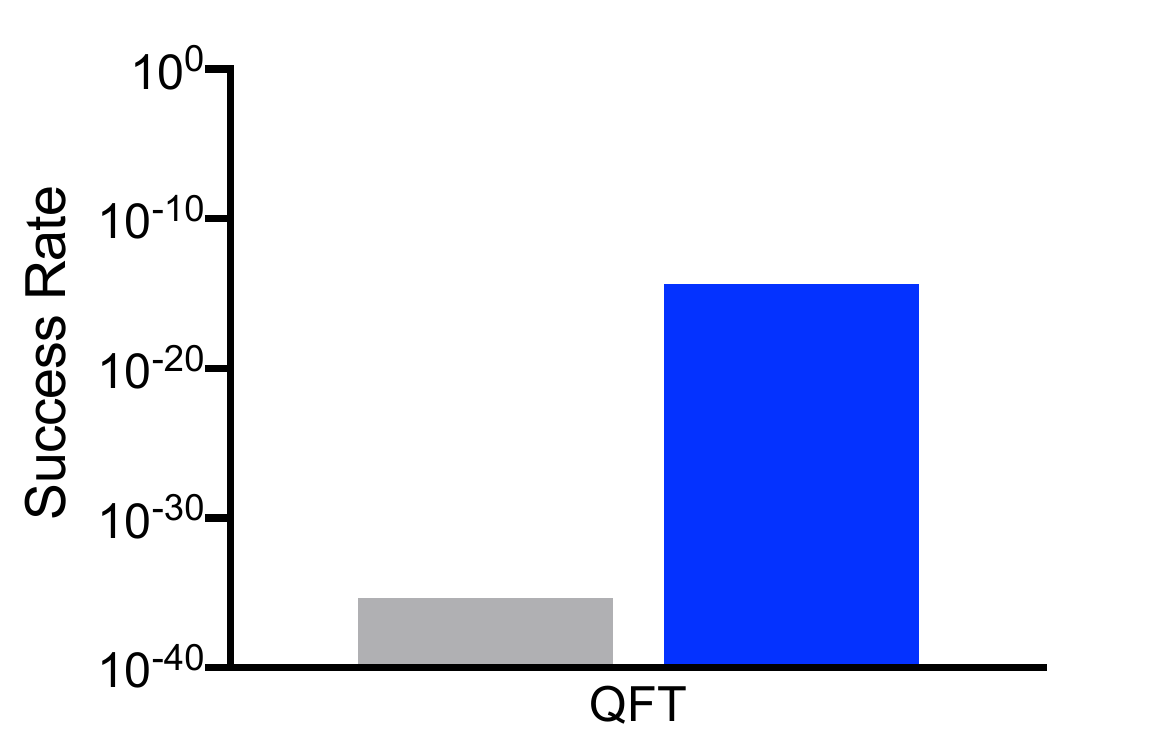}
\label{fig:bsr_qft}
}
\hspace{-2mm}
\subfigure[Success rate (SQRT)]
{
\includegraphics[width=0.23\textwidth,keepaspectratio]{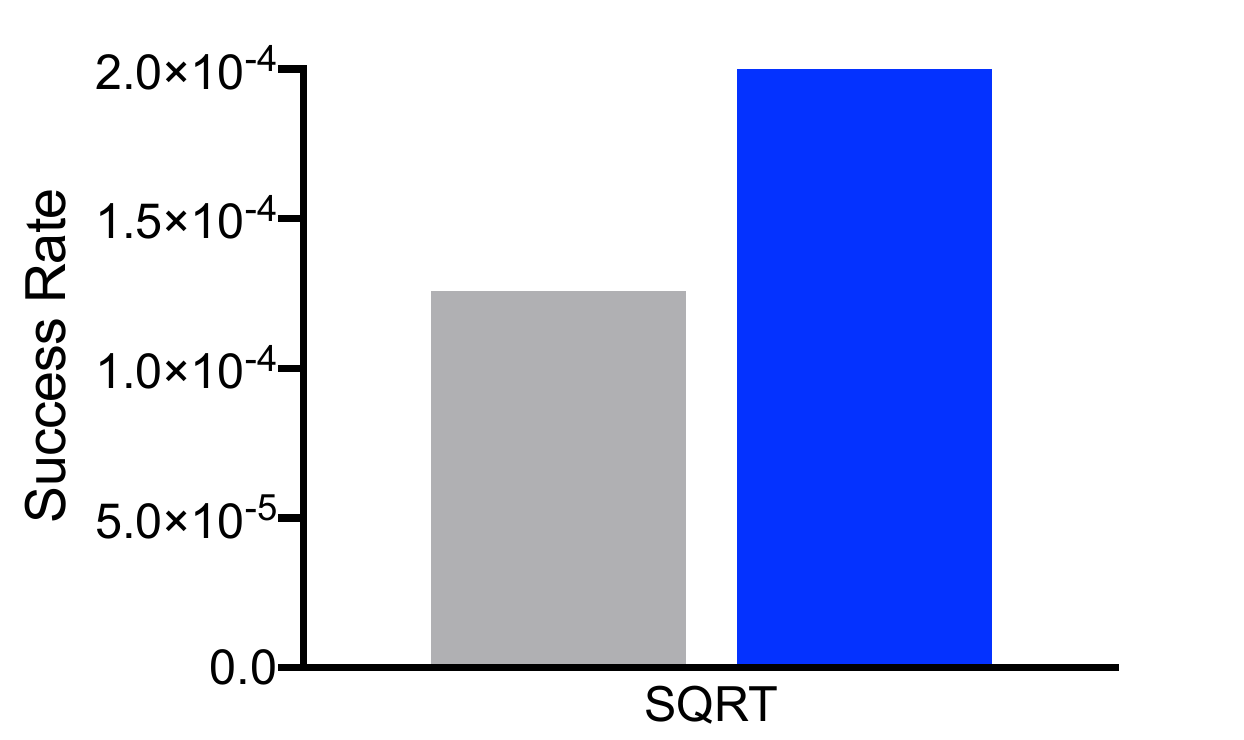}
\label{fig:bsr_sqrt}
}
\caption{Comparing LinQ swap insertion with the baseline swap insertion.}
\label{fig:base}
\end{figure}

We demonstrate the results with laser head size of 16. Figure~\ref{fig:bo} shows the opposing swap ratio, and Figure~\ref{fig:bs} shows swap counts with baseline and LinQ swap insertion. Our LinQ heuristic search reduces the total number of swaps and also increases the opposing swap ratio. For BV applications, however, LinQ does not create any opposing swaps because of the BV circuit structure \cite{bernstein1997quantum,wright2019benchmarking}. 

Figure~\ref{fig:bm} shows the number of tape moves (lower is better). Compared with the baseline, LinQ can synthesize circuits with fewer swaps and consequently fewer gates in total. As a result, fewer tape moves are required for the circuit execution. The success rate is related to the total number of gates and tape moves. Fewer gates and tape moves results in a higher success rate. Thus, LinQ can also achieve higher success rates, as shown in Figure~\ref{fig:bsr_bv} - \ref{fig:bsr_sqrt}.

\begin{figure*}[!t] \centering
\includegraphics[width=0.32\textwidth,keepaspectratio]{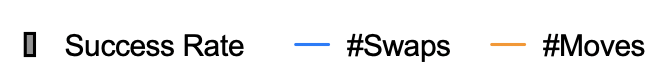}
\\
\subfigure[BV]
{
\includegraphics[width=0.3\textwidth,keepaspectratio]{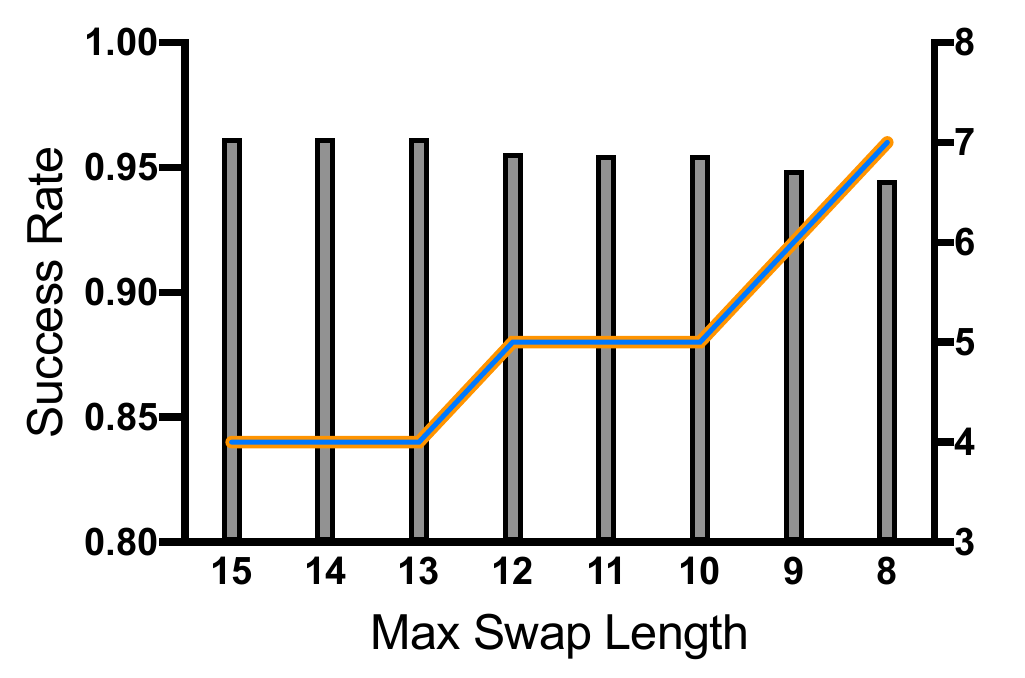}
\label{fig:sl_bv}
}
\hspace{-2mm}
\subfigure[QFT]
{
\includegraphics[width=0.3\textwidth,keepaspectratio]{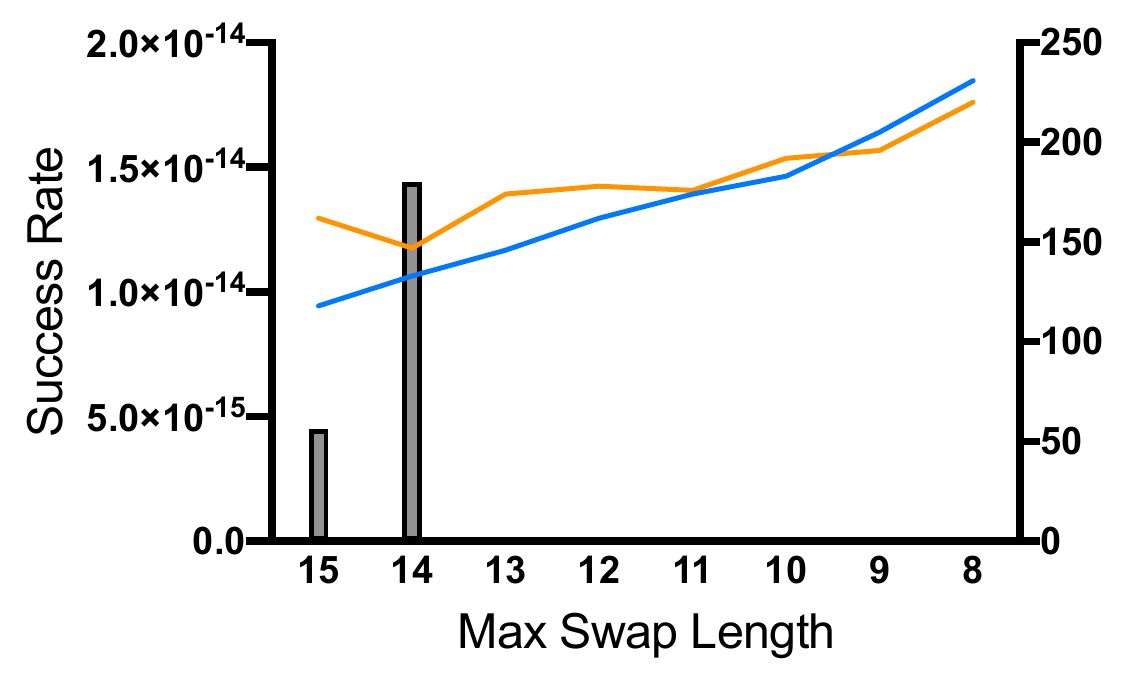}
\label{fig:sl_qft}
}
\hspace{-2mm}
\subfigure[SQRT]
{
\includegraphics[width=0.3\textwidth,keepaspectratio]{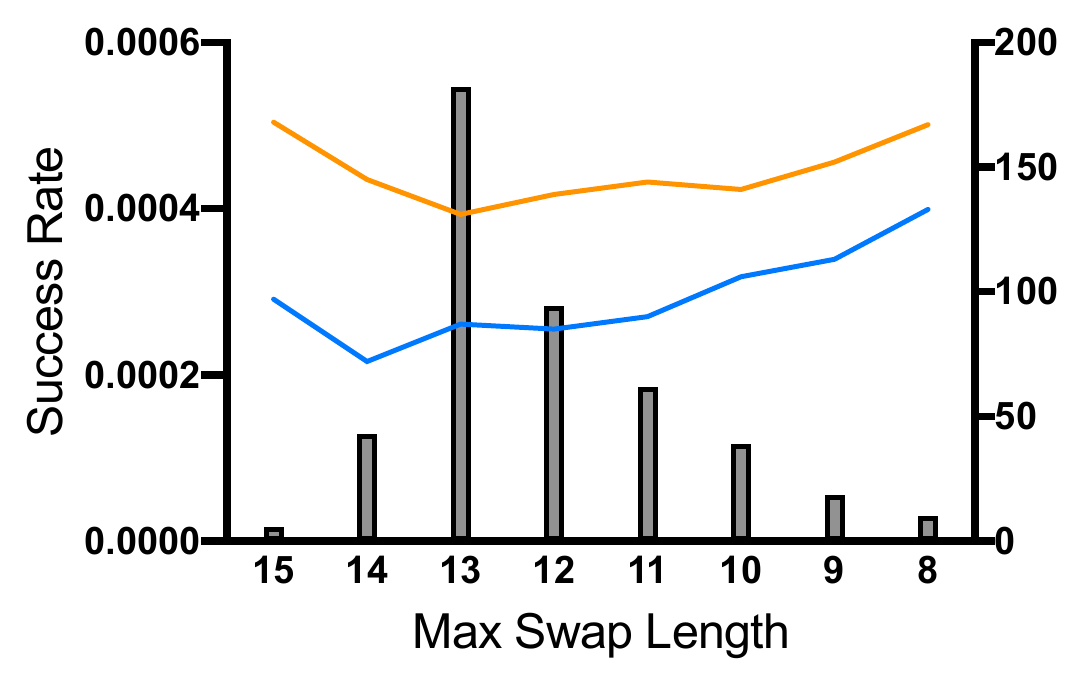}
\label{fig:sl_sqrt}
}
\caption{Success rates under different $MaxSwapLen$ restriction. For larger swap length, the circuit may need more tape moves. If the swap length is too short, however, the required number of tape moves increases as the total number of gates is increased. There is a swap length sweet spot depending on applications. For BV, the swap counts and move counts are the same so the lines are overlapped.}
\label{fig:swap_length}
\end{figure*}

Figure~\ref{fig:swap_length} demonstrates the effect of limiting swap length. Here we show the results of benchmarks on the devices with tape head size of 16. With LinQ swap insertion, while the swap is restricted to a shorter distance, the swap gate count could be slightly increased; but this restriction can potentially reduce the number of tape moves, leading to an improved success rate. 
Figure~\ref{fig:sl_bv} shows the results of BV. The success rates are almost the same for the $MaxSwapLen$ of 15 through 13, because there is no difference in the number of swaps and tape moves. While the max swap length is limited to a lower number, the number of swap gates and moves are increased, and the success rate drops. Figure~\ref{fig:sl_qft} shows the number of swaps and tape moves for QFT under different swap length restrictions. When the maximal swap length is 14, the number of tape moves is the lowest. Although the swap count is increased compared with $MaxSwapLen = 15$, the overall success rate is increased. Therefore, we can get the highest success rate when the maximal swap length is shortened to 14. For SQRT, shown in Figure~\ref{fig:sl_sqrt}, when $MaxSwapLen = 13$, the swap count is slightly increased, but the tape moves can be significantly reduced. Hence, this application reaches the highest success rate at this configuration. For different applications, the best maximal swap length varies. We can iterate the LinQ procedure to find the best choice.

\subsection{Architecture Comparison}\label{sec:eval_arch}
In this section, we compare the \tilt{} architecture with head sizes of 16 and 32 with ideal trapped-ion (Ideal TI) devices and the QCCD system presented in~\cite{murali2020architecting}. 

\noindent\textbf{Ideal TI.} An ideal trapped-ion device means that the device has enough laser controls for each qubit. Consequently, two-qubit gates can be performed on an arbitrary pair of qubits. As a result ion shuttling and swap gate insertion are unnecessary under our error model. With comparing to this ideal case, we can learn how close we are to the optimal solution.

\noindent\textbf{QCCD.} A recent study on QCCD is presented in~\cite{murali2020architecting}. In our evaluation, we apply the same noise model and topology evaluated in the study. The QCCD configurations are 15-35 ions per trap in a linear topology, and we select the highest reported fidelity with AM gates for the comparison.

\begin{figure*}[!t]
\centering
\subfigure[Success rates of applications (ADDER, BV, QAOA, and RCS).]
{
\includegraphics[width=0.45\textwidth,keepaspectratio]{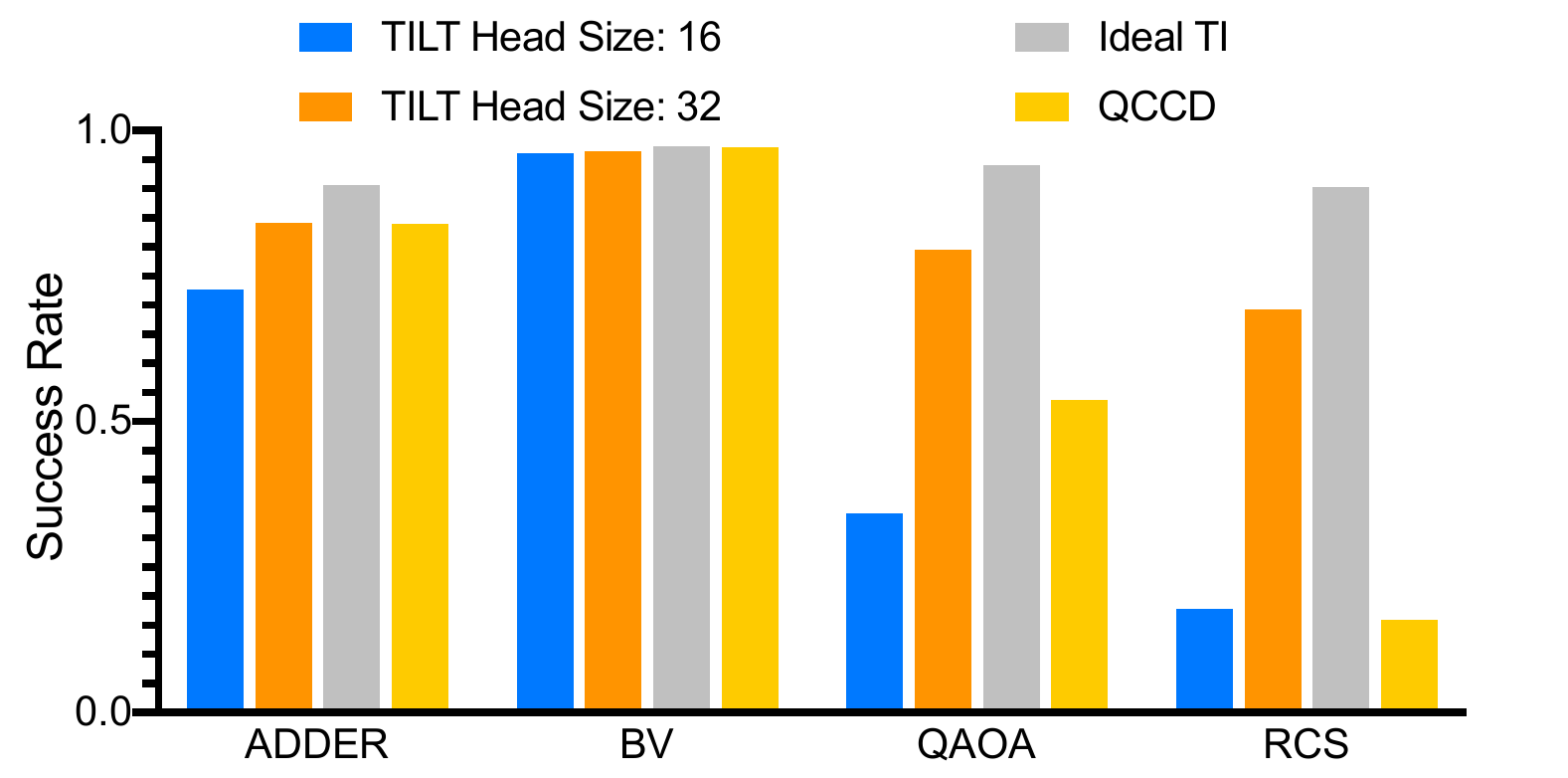}
\label{fig:sc_sr_1}
}
\hspace{-0mm}
\subfigure[Success rates of applications (QFT and SQRT).]
{
\includegraphics[width=0.45\textwidth,keepaspectratio]{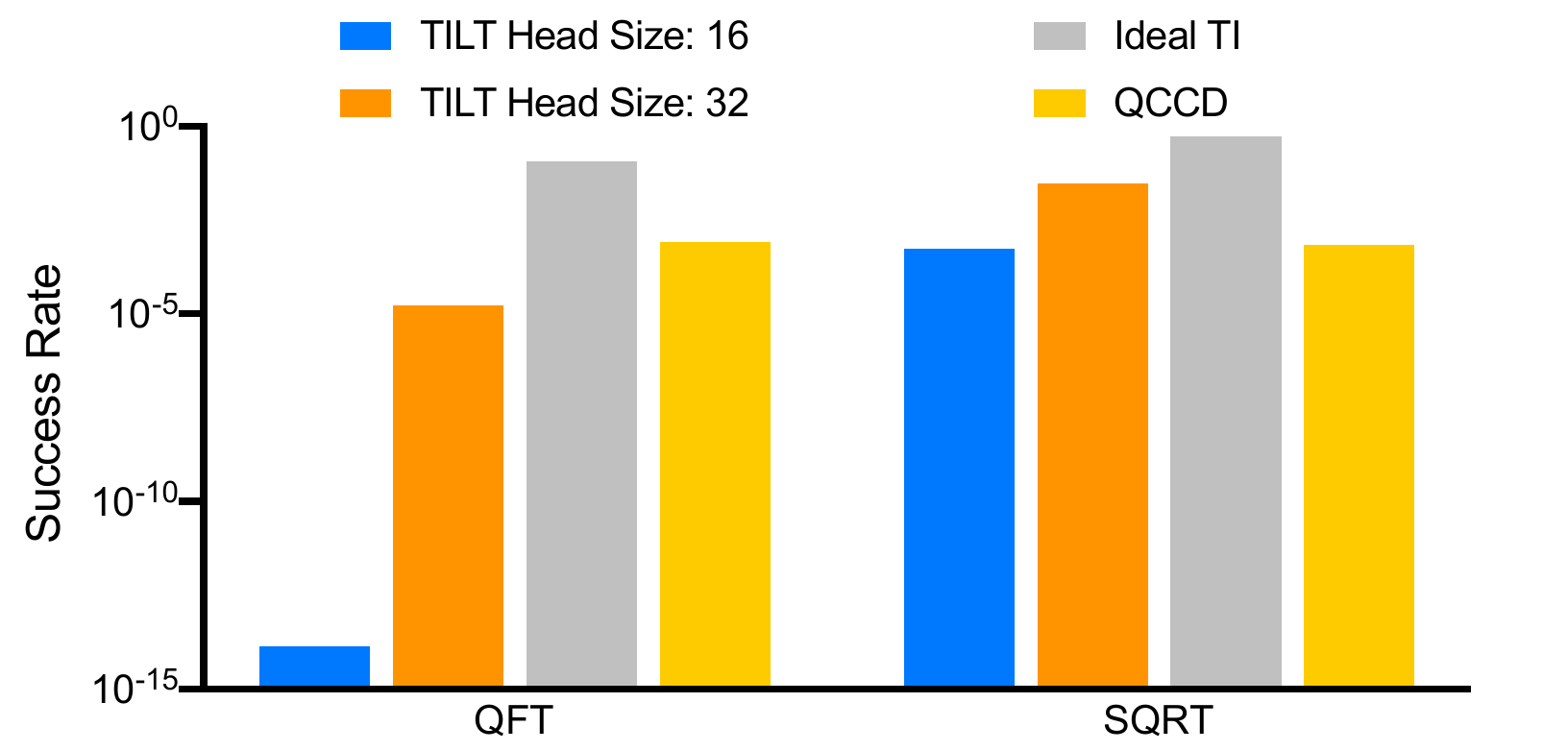}
\label{fig:sc_sr_2}
}
\caption{Success rates on different device configurations (higher is better).}
\label{fig:sc_sr}
\end{figure*}

Figure~\ref{fig:sc_sr} shows the results from our simulation. The success rate of \tilt{} is related to the tape head size. For different tape head sizes, a larger tape head can cover a wider range of execution zone, reducing the number of swaps as well as tape moves and achieving a higher success rate. For ADDER and BV benchmarks, \tilt{} has the same performance as QCCD. The reason is that the circuits are primarily operating on a few qubits. For QCCD, only a few shuttles have to be performed. Similarly for \tilt{}, only a few tape moves are required to perform the circuit execution. However, for QAOA and RCS applications, the success rates of \tilt{} are significantly higher than QCCD. As discussed in Section~\ref{sec:compare}, frequent short-distance two-qubit gates will cause QCCD to perform split/merge and shuttling operations periodically if the involved qubits are not in the same trap, while \tilt{} can possibly schedule those operations in one tape move. Since QAOA and RCS repeat the operations on the pair of short-distance qubits, \tilt{} outperforms QCCD in these results. For the circuits composed of short- and long-distance two-qubit gates (QFT and SQRT), \tilt{} has lower success rate on QFT. This is as expected because the long-distance operations require multiple swaps and tape moves for \tilt{}. 

Our evaluation results suggest that \tilt{}'s success rate is better than that of QCCD for applications with a majority of the communication occurring within the width of a tape head.

\subsection{Compilation Time and Execution Time}
Table~\ref{tab:results} shows the compilation results. The column labels $t_{swap}$ and $t_{move}$ are the compilation times for swap insertion and tape move scheduling for the benchmarks using our LinQ toolflow. The $t_{swap}$ is short for each application. As discussed in Section~\ref{sec:linq_swap}, the worst-case time complexity of searching for swap insertion is $O(NG)$. As a result, this heuristic search is scalable for larger applications. For tape move scheduling, as discussed in Section~\ref{sec:linq_move}, the worst-case scheduling time is $O((N-L)DL)$. Since the circuit depth is increased when more swap gates are inserted, the scheduling time for QFT and SQRT is longer than for other applications. The number of swaps can be reduced by using a larger tape head, leading to lower circuit depth and a reduced $t_{move}$. The results and complexity analysis suggest that LinQ is applicable to larger circuits.

\begin{table*}[h!]
\centering
\caption{LinQ compilation results. }
\small
\begin{tabular}{l|ccccc | ccccc}
\hline\hline
\multirow{2}{*}{Application} & \multicolumn{5}{c|}{Head Size: 16}&\multicolumn{5}{c}{Head Size: 32}\\
& $t_{swap}$(s) & $t_{move}$(s) & \#moves & dist($\mu m$) & $t_{exec}$(s) &$t_{swap}$(s) & $t_{move}$(s) & \#moves & dist($\mu m$) & $t_{exec}$(s) \\
\hline
ADDER&0.002&6.214&10&104&2.967&0.002&3.248&5&68&3.252\\
BV&0.004&1.575&4&49&0.856&0.003&0.899&2&33&0.987\\
QAOA&0.011&2.750&18&232&1.564&0.011&1.101&4&72&1.357\\
RCS&0.001&3.148&65&992&1.704&0.001&0.680&11&214&0.856\\
QFT&9.206&55.617&162&2002&24.820&6.135&31.206&69&1276&33.876\\
SQRT&1.344&129.901&168&1816&46.554&1.075&55.234&76&1068&40.817\\
\hline\hline
\end{tabular}

\small{$t_{swap}$: compile time for swap insertion. $t_{move}$: compile time for tape move scheduling. \textbf{\#moves}: total number of tape moves. \\\textbf{dist}: total tape move distance. $t_{exec}$: program execution time.}
\label{tab:results}
\end{table*}


To estimate the program execution time, we analyze the compiled circuit depth and tape move distance. We model the gate time as shown in Equation~\ref{eq:gate_time}, and the shuttling rate ($t_m$) as $1 \mu m/\mu s$~\cite{bruzewicz2019trapped}. We obtain the program execution time as following, 
\begin{equation}\label{eq:t_exe}
t_{exe} = t_m\times dist + \sum_{d} t_{d},
\end{equation}
where $t_d$ is the maximum gate time for the $d$-th depth. From our simulation, the applications can be finished within seconds. The results show that \tilt{} machines are viable for running large applications, like our benchmarks, since trapped-ion qubits have long coherence times~\cite{timoney2011quantum}.

\section{Trapped-Ion Scaling}\label{sec:scaling}
Since ions are fundamentally identical and the same trapping parameters can handle a range of ion counts, scaling qubit count is not as significant a barrier as in other architectures. As we increase the number of ions, however, the heating cost of shuttling will also increase. This will lead to lower gate reliability, so there is an effective limitation on the number of ions in a single trap. 

Various trapped-ion scaling techniques can be combined with \tilt{}. In this section we discuss some additional techniques and technologies which could be used in conjunction with \tilt{} to further scale trapped ion quantum computers.

\noindent\textbf{Sympathetic Cooling.} Gate reliability depends on the thermal energy in the chain. Sympathetic cooling is a technique to mitigate thermal heating, compensating for the increased heating rates expected in large-scale trapped-ion devices~\cite{mudrich2002sympathetic,bruzewicz2019trapped}. A dual-species ion chain is composed of two different types of ions, one for storing information and which can be cooled by another laser-cooled ion during circuit execution without damaging the stored information. \tilt{} architectures are compatible with sympathetic cooling techniques, which would reduce the heating due to shuttling and allow for longer circuits. With cooling ions, a few modifications would be made for LinQ toolflow to consider the tape scheduling under the head of cooling laser beams. However, the fundamental design goal is still the same, minimizing the number of tape moves.

\noindent\textbf{QCCD Architectures.} The QCCD architecture is a well-known design for constructing a large-scale quantum computer~\cite{qccd1, murali2020architecting}. QCCD systems have multiple small traps that are interconnected by segments and junctions. Before shuttling an ion from one trap to another, the ion must be split from the source chain. The split ion then is shuttled to the target trap and merged into the destination chain. 

The operations necessary for individual ion shuttling have high costs in terms of shuttling complexity. Individual ion shuttling requires expensive split/merge and junction crossing maneuvers, which lead to more thermal energy entering the system. QCCD architectures could be combined with \tilt{} as a combined architecture where the \tilt{} systems discussed in this work would be a primitive for constructing the QCCD. Trap capacities could become larger, which might be useful for applications in which the circuit naturally breaks up into larger densely-communicating chunks.

Such an architecture could combine the strengths of each system, allowing for circuits which have significant medium to long-range communication to live within a single trap, while also allowing other sections of the circuit which might not require such distant communication to occur within smaller higher-fidelity trapping zones. Heterogeneous systems like this, where different traps serve different purposes, would be a logical evolution of trapped ion quantum computers into a more modular framework similar to modern classical computers.

\noindent\textbf{Photonic Interconnects.} \tilt{} can also be utilized in modular quantum computer architectures such as the modular universal scalable ion-trap quantum computer (MUSIQC) architecture proposed in \cite{kim2011modular, monroe2014large}. The element logic units (ELUs) in this architecture consist of arrays of ions, and one or more ions in each ELU are coupled to photonic quantum channels through photonic interconnects. With these interconnects, we can use \tilt{} architectures as the fundamental building blocks of ELUs to achieve modular \tilt{} architectures and further scale QC devices, similar to the QCCD discussion above.
\section{Related Work}\label{sec:related}

Several studies of  qubit mapping and swap insertion problems have been carried out for superconducting systems. One common approach is to formulate the problems in mathematical form, such as integer linear programming, and then utilize software solvers to find the optimal solutions~\cite{bahreini2015minlp,wille2014optimal,lye2015determining,maslov2008quantum,wu2019ilp}. This method is guaranteed to provide an optimal solution. However, since the time to solution grows exponentially with the number of qubits and gates, so scaling this approach to large NISQ programs is infeasible. In \cite{wu2019ilp}, a circuit partitioning approach is proposed. However, There is a large amount of wasted tape movement between each circuit block. Thus, our scheduling algorithm outperforms the ILP-based approach. Another approach is to apply dynamic programming to find the optimal solutions~\cite{siraichi2018collange,itoko2020optimization}. However, this method only works for circuits with 10 or fewer qubits since the time scaling is exponential. To avoid long execution times, several researchers have chosen to apply heuristic search algorithms~\cite{itoko2020optimization, li2019tackling,zulehner2018efficient,alfailakawi2014lnn,saeedi2011synthesis,wille2016look,kole2016heuristic,bhattacharjee2018novel}. Overall, most of the previous studies focus on compilation techniques for superconducting systems. In our work, we develop a scalable software toolflow to compile high-level quantum programs for \tilt{} architecture, a novel trapped-ion system.

Previous studies have evaluated the performance of real devices with less than 20 qubits. In one case a fully connected 5-qubit trapped-ion system is compared with a 5-qubit superconducting transmon system~\cite{linke2017experimental}. In another study, several NISQ benchmarks are performed, comparing the trapped-ion system with superconducting systems~\cite{murali2019full}. These studies show that trapped-ion systems provide high program success rates because of the dense connectivity and higher gate fidelities compared with superconducting systems. In our work, we compare 64-qubit \tilt{} systems to QCCD systems using our simulation tools and published results.

Several studies have proposed different strategies to scale trapped-ion quantum computers. Sargaran et al. proposed the SAQIP to apply reconfigurable optical interconnects for scalable trapped-ion systems~\cite{sargaran2019saqip}. Similarly, \cite{kim2011modular, monroe2014large} proposed the MUSIQC architecture that uses photonic interconnects to link modular elementary logic units. This architecture can scale to thousands of qubits and support fault-tolerant error correction. Lekitsch et al.~\cite{lekitsch2017blueprint} proposed a blueprint for scalable trapped-ion systems that use microwave-based quantum gates with on-chip control electronics to control an arbitrary number of qubits. While these studies focus on very large systems that are unlikely to be built in the next decade, \tilt{} architectures provide a practical near-term implementation to scale trapped-ion systems.

A recent study focuses on 50--100 qubit scale modular QCCD-based trapped-ion devices~\cite{murali2020architecting}. They propose the use of simulation techniques to study the impact of trap sizes, topology, and gate implementations. In our work, we provide simulation techniques to study \tilt{} architectures. Architectural simulations allow us to evaluate the performance of different approaches of scaling trapped-ion systems before building them.

\section{Conclusion}\label{sec:conc}
Trapped-ion technologies are a promising implementation for building practical quantum computers. Since all ions are identical, ions can be added to a long chain in order to scale the QC device. The \tilt{} architecture is able to avoid issues with addressability as the number of ions addressed is held constant independent of chain length. Shuttling a small chain of ions has been demonstrated in \cite{fallek2016transport}, and larger \tilt{} demonstrations are planned \cite{staq2020}. In this work, we show that \tilt{} architectures offer a viable path toward QC devices approaching 60+ qubits. We present our optimizing compiler and simulator for \tilt{} architectures. With a gate fidelity model derived from real experiments, our compiler performs qubit mapping, swap insertion, and tape move scheduling for NISQ applications within a few minutes. The results suggest that TILT can outperform QCCD in a range of NISQ applications. We also discuss using TILT as a building block to extend other scalable trapped-ion quantum computing proposals. Our evaluation of the TILT architecture offers insights for future architecture design. 
\section*{Acknowledgements}
This work is funded in part by EPiQC, an NSF Expedition in Computing, under grants CCF-1730449/1832377; in part by STAQ, under grant NSF Phy-1818914; in part by NSF-OMA-2016136; and in part by DOE grants DE-SC0020289 and DE-SC0020331. This material is based upon work supported by the U.S. Department of Energy, Office of Science and Technology, under contract DE-AC02-06CH11357. DMD is also funded in part by an NSF QISE-NET fellowship (1747426). We thank Prakash Murali for fruitful discussions. We also thank the anonymous reviewers for their valuable comments and suggestions.


\bibliographystyle{IEEEtranS}
\bibliography{refs}

\end{document}